\documentclass[12pt]{article}

\usepackage{cite}
\usepackage{subfigure}
\usepackage{multirow}
\usepackage{helvet}
\usepackage{amsmath}
\usepackage{amssymb}
\usepackage{setspace}
\usepackage{setspace}
\usepackage[dvips]{graphicx}
\usepackage{epsfig}

\begin{document}

\titlepage                                                    
\begin{flushright}                                                    
IPPP/13/102  \\
DCPT/13/204 \\                                                                                                       
\end{flushright} 
\vspace*{0.5cm}
\begin{center}                                                    
{\Large \bf Modeling exclusive meson pair production at hadron colliders}\\


\vspace*{1cm}
                                                   
L.A. Harland--Lang$^{1}$, V.A. Khoze$^{1,2}$, M.G. Ryskin$^{2}$ \\                                                 
                                                   
\vspace*{0.5cm}                                                    
${}^1$ Institute for Particle Physics Phenomenology, University of Durham, Durham, DH1 3LE          \\
${}^2$ Petersburg Nuclear Physics Institute, NRC Kurchatov Institute, Gatchina, \linebreak[4]St. Petersburg, 188300, Russia                                              
                                                    
\vspace*{1cm}                                                    
                                                    
\begin{abstract}                                                    
\noindent

We present a study of the central exclusive production of light meson pairs, concentrating on the region of lower invariant masses of the central system and/or meson transverse momentum, where perturbative QCD cannot be reliably applied. We describe in detail a phenomenological model, using the tools of Regge theory, that may be applied with some success in this regime, and we present the new, publicly available, \texttt{Dime} Monte Carlo (MC) implementation of this for $\pi\pi$, $KK$ and $\rho\rho$ production. The MC includes a fully differential treatment of the survival factor, which in general depends on all kinematic variables, as well as allowing for the so far reasonably unconstrained model parameters to be set by the user. We present predictions for the Tevatron and LHC, discuss and estimate the size of the proton dissociative background, and show how future measurements may further test this Regge--based approach, as well as the soft hadronic model required to calculate the survival factor, in 
particular in the presence of tagged protons.

\end{abstract}                                                        
\vspace*{0.5cm}                                                    
                                                    
\end{center}  

\section{Introduction}

The study of central exclusive production (CEP) processes of the type
\begin{equation}\label{exc}
pp({\bar p}) \to p+X+p({\bar p})\;,
\end{equation}
can significantly extend the physics programme at high--energy hadron colliders. Here $X$ represents a system of invariant mass $M_X$, and the `$+$' signs denote the presence of large rapidity gaps. Such reactions provide a very promising way to investigate both QCD dynamics and new physics in hadron collisions, with there recently being a renewal of interest in the CEP process, see for example~\cite{Martin:2009ku,Albrow:2010yb,HarlandLang:2013jf}  for reviews and further references.

One particularly interesting example of such processes is the exclusive production of light meson pairs (i.e. $X=M_3 M_4=\pi\pi, KK, \rho\rho, \eta(')\eta(')$...). This has been the focus of recent studies in~\cite{HarlandLang:2011qd,Harland-Lang:2013ncy,Harland-Lang:2013qia}, where a perturbative approach, combining the pQCD based `Durham model' of the CEP process (see e.g.~\cite{HarlandLang:2010ep} and references therein) with the `hard exclusive' formalism described in~\cite{Brodsky:1981rp} (see also~\cite{Benayoun:1989ng}) to calculate the $gg \to M_3 M_4$ amplitude, was applied. As summarised in~\cite{Harland-Lang:2013qia}, such an approach, which should be valid at sufficiently high meson transverse momentum $k_\perp$ (and/or pair invariant mass $M_X$), leads to some very non--trivial phenomenological predictions, while the corresponding $gg \to M_3 M_4$ helicity amplitudes have some remarkable theoretical features. 

However, the study of meson pair CEP in fact has a long history, which far predates this approach~\cite{Kaidalov:1974qi,Azimov:1974fa,Pumplin:1976dm,Desai:1978rh} (see also~\cite{HarlandLang:2012qz,Lebiedowicz:2009pj,Lebiedowicz:2012nk} for references and recent studies). In these cases, the production process was instead considered within the framework of Regge theory (see for example~\cite{Collins:1977jyp} for an introduction), with the meson pair produced by the exchange of two Pomerons in the $t$--channel, as shown in Fig.~\ref{npip}. Such a `non--perturbative' picture should be relevant at lower values of the meson transverse momentum $k_\perp$, where the cross sections are largest, and may be particularly important for the case of flavour--non--singlet mesons ($\pi\pi$, $KK$...), for which the perturbative contribution is expected to be dynamically suppressed, see~\cite{HarlandLang:2011qd}. At lower meson invariant masses $M_X \lesssim 2$ GeV there will also in general be a host of different resonances 
which lie on top of, and interfere with, this continuum contribution; the production of lower mass resonances was recently examined in for example~\cite{Lebiedowicz:2013ika}, while in~\cite{HarlandLang:2012qz}  the continuum background to the production of the higher mass $\chi_{c(0,2)}$ states via  two--body $\pi^+\pi^-, K^+K^-$ decays was considered. Moreover, we may expect data on the CEP of meson pairs to be forthcoming from CMS~\cite{enterria}, CMS+Totem~\cite{CMSeds,Oljemark:2013wsa,Oljemarkeds}, ATLAS+ALFA~\cite{Staszewski:2011bg,Sykoraeds}, RHIC~\cite{Leszek}, and LHCb~\cite{Paula}, while the results of the new analysis of the CDF data at $\sqrt{s} = 900$ and 1960 GeV have been presented in~\cite{Albrow:2013mva,Mikeeds}. 

For these reasons it is important to give a careful theoretical consideration to meson pair production in this experimentally most accessible regime. In this paper we consider a phenomenological model, outlined in~\cite{HarlandLang:2012qz}, and show how the undetermined aspects of such a model can be constrained by the forthcoming and existing data, which in this way can serve as a probe of such a Regge--based framework. We will also show how the observation of CEP processes, such as meson pair production, in the presence of tagged protons, can act as a very sensitive test of the models of soft diffraction, see e.g.~\cite{Ryskin:2009tk,Khoze:2013dha,Khoze:2013jsa} and~\cite{Gotsman:2012rm,Gotsman:2012rq}, which are needed to calculate the `survival probability' for no additional soft rescattering between the colliding protons, as well as providing a description of other hadronic (total, elastic, diffractive) scattering data. In particular, a measurement of the distribution in azimuthal angle between the 
outgoing intact protons can provide a fully differential test of the soft survival factors. Such measurements are under consideration at the LHC, with the CMS+Totem~\cite{CMSeds,Oljemark:2013wsa,Oljemarkeds} and ATLAS+ALFA~\cite{Staszewski:2011bg,Sykoraeds} detectors, in particular during special low luminosity running conditions, and are already being made at RHIC by the STAR collaboration~\cite{Leszek} and by the COMPASS fixed--target experiment at CERN~\cite{fortheCOMPASS:2013vda}.

Motivated by this, we present in this paper the new public \texttt{Dime} Monte Carlo~\cite{dime} for meson pair ($\pi\pi, KK, \rho\rho$) CEP via this non--perturbative mechanism. We give the user freedom to set the most important aspects and parameters of the model, so that these can be compared with and adjusted to future data. We also include the soft survival factor at the fully differential level, which (as described in e.g.~\cite{HarlandLang:2010ep}) is crucial to give a complete prediction, in particular when considering the kinematic distributions of the outgoing protons.

Finally, we note that, with the exception of the dedicated TOTEM + CMS measurements~\cite{CMSeds,Oljemark:2013wsa,Oljemarkeds}, the LHC experiments have so far studied CEP processes without the use of forward spectrometers to tag the outgoing intact protons, instead applying a Large Rapidity Gap (LRG) veto on additional particles in a certain rapidity region. Although this can be used to select a reasonable fraction of purely exclusive events, in general certain regions in the forward/backward directions are uninstrumented, and so there is always a possibility of an admixture of events in which the proton dissociates into a hadronic system, with the secondaries from this system not seen in the detector. This contribution will lead to a larger measured cross section than the theoretical predictions for CEP, which assume that the outgoing protons remain intact. This contamination may be particularly important in the case that some suppression is expected in the purely exclusive cross section, due to for 
example the $J_z^{P}=0^+$ quantum number selection rule (see~\cite{HarlandLang:2012qz} for some discussion of this in the case of $\chi_{c2}$ production). For this reason, we we will also discuss and present estimates in this paper for the expected size of the proton dissociative contamination in events selected experimentally with a large rapidity gap veto.

The outline of this paper is as follows. In Section~\ref{theory} we describe the different aspects of the model, considering the choice of form factor for the meson--Pomeron coupling in Section~\ref{off}, the possibility of additional particle production in the Pomeron fusion subprocess in Section~\ref{add}, Reggeization of the $t$--channel meson exchange in Section~\ref{secreg}, secondary Reggeon contributions in Section~\ref{secregs} and soft survival effects in Section~\ref{secsurv}. In Section~\ref{secdime} we describe in more detail the \texttt{Dime} MC, which implements the model described in the preceding sections. In Section~\ref{numer} we present a selection of numerical predictions, and discuss the possibilities for measurements at hadron colliders. In Section~\ref{secdiss} we discuss the issue of proton dissociation, as described above. Finally, in Section~\ref{conc} we conclude.

\section{Theory}\label{theory}

\begin{figure}[h]
\begin{center}
\includegraphics[scale=1.0]{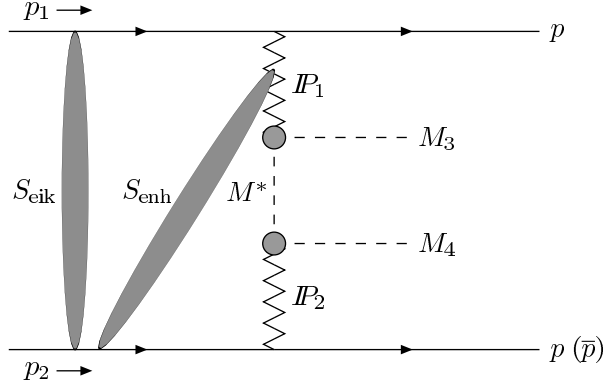}
\caption{Representative diagram for the non-perturbative meson pair ($M_3$, $M_4$) CEP mechanism, where $M^*$ is an intermediate off-shell meson of type $M$. Eikonal and (an example of) enhanced screening effects are indicated by the shaded areas.}\label{npip}
\end{center}
\end{figure}

The model we will consider in this paper applies the well--established tools of Regge theory~\cite{Collins:1977jyp}, and is represented in Fig.~\ref{npip}. In this `one--meson exchange' model (see for instance~\cite{Azimov:1974fa,Pumplin:1976dm,Lebiedowicz:2011nb}) the mesons are produced via Pomeron--Pomeron fusion, with an intermediate off--shell meson exchanged in the $t$--channel. The CEP cross section is given by
\begin{equation}\label{ncross}
\sigma^{CEP}=\frac {1}{16\pi(16\pi^2)^2}\int dp^2_{1\perp}dp^2_{2\perp}dy_3dy_4dk^2_{\perp}\frac{|\mathcal{M}|^2}{s^2}\;,
\end{equation}
where $\sqrt{s}$ is the c.m.s. energy, $p_{1\perp}, p_{2\perp}$ are transverse momenta of the outgoing protons, $k_\perp$ is the meson transverse momentum and $y_{3,4}$ are the meson rapidities. Ignoring secondary Reggeon contributions and soft survival effects for simplicity (these will be discussed in the sections which follow), the production amplitude, $\mathcal{M}$, is given by the sum $\mathcal{M}=\mathcal{M}_{\hat{t}}+\mathcal{M}_{\hat{u}}$ of the $t$ and $u$--channel contributions, with $\hat{t}=(P_1-k_3)^2$, $\hat{u}=(P_1-k_4)^2$, where $P_i$ is the momentum transfer through Pomeron $i$, and $k_{3,4}$ are the meson momenta. We have
\begin{equation}\label{namp}
\mathcal{M}_{\hat{t}}=\frac 1{M^2-\hat{t}} F_p(p^2_{1\perp})F_p(p^2_{2\perp})F^2_M(\hat{t})\sigma_0^2
\bigg(\frac{\hat{s}_{13}}{s_0}\bigg)^{\alpha_{I\!\!P}(p^2_{1\perp})}\bigg(\frac{\hat{s}_{24}}{s_0}\bigg)^{\alpha_{I\!\!P}(p^2_{2\perp})}\;,
\end{equation}
where $M$ is the meson mass, we take $s_0=1\,{\rm GeV}^2$ and $\alpha_{I\!\!P}(p^2_{i\perp})=1.08-0.25\, p^2_{i\perp}$, for $p^2_{i\perp}$ measured in ${\rm GeV}^2$~\cite{Donnachie:1992ny}, and $s_{ij}=(p_i'+k_j)^2$ is the c.m.s. energy squared of the final--state proton--meson system $(ij)$.  The proton form factors are usually taken for simplicity to have an exponential form, $F_p(t_i)=\exp(B_it_i/2)$, as was assumed in the previous work of~\cite{HarlandLang:2012qz}, however as we shall see, the latest fit of~\cite{Khoze:2013dha} for the soft survival factor suggests a different parameterization should be taken, see Section~\ref{secsurv}. We can see from (\ref{namp}) that the cross section normalisation is set by the total meson--proton cross section $\sigma_{\rm tot}(M p)=\sigma_0 (s_{ij}/s_0)^{\alpha(0)-1}$ at the relevant sub--energy; the factor $\sigma_0$ can be extracted for example from the fits of~\cite{Donnachie:1992ny}. While this is therefore well constrained for the cases of $\pi\pi$ and $KK$ 
production, there remain other elements and possible additions to the model, which as we shall see are in general quite poorly constrained by the relatively limited available ISR data. These are: the form factor $F_M(\hat{t})$ in (\ref{namp}) of the Pomeron coupling to the off--shell meson, the possibility to produce additional particles in the Pomeron fusion subprocess, and the effect of Reggeization of the meson exchange in the $t$--channel. We will consider each of these in turn, before discussing the inclusion of secondary Reggeons and soft survival effects.

\subsection{Off--shell meson form factor}\label{off}

The $F_M(\hat{t})$ in (\ref{namp}) is the form factor for the coupling of the Pomeron to the outgoing meson and the off--shell $t$--channel meson exchange. Unfortunately no direct measurement of this form factor for light mesons exists: we will see that data from the ISR place some constraint but nonetheless the shape of the form factor, in particular at higher values of $|\hat{t}|$, is unknown. We therefore treat it as a phenomenological input in our model, and will consider for concreteness three different possibilities
\begin{align}\label{Fexp}
 F^{\rm exp}_M(\hat{t})&=\exp\left(b_{\rm exp} \hat{t'}\right)\;,\\ \label{Forexp}
 F^{\rm or}_M(\hat{t})&=\exp(-b_{\rm or} \sqrt{-\hat{t'}+a_{\rm or}^2}+a_{\rm or}b_{\rm or})\;,\\ \label{Fpow}
 F^{\rm pow}_M(\hat{t})&=\frac{1}{1-\hat{t'}/b_{\rm pow}}\;,
\end{align}
where $\hat{t'}=\hat{t}-M^2$, and $M$ is the meson mass. That is, a typical `soft' exponential, a power--like behaviour, and a form of the type proposed by Orear~\cite{Orear:1964zz}. Such `Orear--like' behaviour $d\sigma_{el}(pp)/dt\propto \exp(-b\sqrt{-t})$ was observed experimentally~\cite{Orear:1964zz} in the case of proton--proton scattering at relatively large $|t|$, and so we might expect to see a similar type of behaviour here. Theoretically, it was shown in~\cite{Anselm1967479} that this behaviour may result from the contribution of a series of diagrams with different numbers of exchanged Pomerons/Reggeons  (i.e. multi--Reggeon cuts). Assuming a `soft' exponential behavior ($\sim \exp(Bt)$) for one--Reggeon exchange we get an $n$--times smaller slope ($\sim \exp(Bt/n)$) for the $n$--Reggeon exchange amplitude, and the sum of these contributions may be described by an Orear--like form factor (see also \cite{Andreev:1967zz}). Thus (\ref{Forexp}) may be 
considered as an `effective' form factor, containing the contribution of a more complicated set of diagrams (i.e. 
with additional exchanges between the outgoing mesons in the $t$--channel) which are not included explicitly in the calculation.  We replace $\sqrt{-t}$ in (\ref{Forexp}) by $\sqrt{-t+a^2_{or}}$ in order to keep the analytic properties of the amplitude, in particular under $t \leftrightarrow s$ crossing.

All form factors are defined so that they reach unity if the squared $4$--momentum transfer is equal to the meson mass squared $M^2$. In Section~\ref{numer} we will show how a comparison to ISR data for exclusive $\pi^+\pi^-$ and $K^+K^-$ production allows some approximate values for the parameters ($b_{\rm exp}$, $b_{\rm or}$, $a_{\rm or}$, $b_{\rm pow}$) to be extracted. However, as we will see in Section~\ref{numer} the choice of form factor leads to dramatically different behaviour at higher $|\hat{t}|$, beyond the region probed by the ISR data. Moreover, we note that these relatively simple phenomenological forms (\ref{Fexp})--(\ref{Fpow}) may not be expected to hold across the entire $|t'|$ range, for example when the meson transverse momentum, $k_\perp$, becomes sufficiently high, and the `perturbative' regime is reached, where the description discussed in~\cite{HarlandLang:2011qd,Harland-Lang:2013ncy,Harland-Lang:2013qia} should be relevant. These therefore represent our best educated guesses for the 
off--shell meson form factor, the validity of which is to be determined by comparison to future collider data.

\subsection{Additional particle production}\label{add}

A correction we may consider to the simple model of (\ref{namp}) is the suppression which comes from the requirement that no additional particles are produced in the meson pair production process, that is due to screening corrections; in terms of the Reggeon formalism these absorptive corrections are described by the exchange of additional (one or more) Pomerons in the diagram shown in Fig.~\ref{npip}. First, there is the exchange between the two incoming (outgoing) protons: this is just the usual `eikonal' survival factor $S_{\rm eik}$, which we will discuss in Section~\ref{secsurv}. In addition to this, we have the possibility of rescattering between the protons and the outgoing mesons. However, as discussed in more detail in~\cite{HarlandLang:2012qz}, such an interaction is either suppressed by the small phenomenological value of the triple--Pomeron coupling in the case of proton--Pomeron rescattering, or by the small size of the produced `half--dressed' mesons $\sim 1/\sqrt{\hat{s}}$ in the case of 
proton--meson rescattering. Such effects will therefore be ignored in what follows, and in particular in the \texttt{Dime} MC.

We must also in principle consider the possibility of additional meson--meson rescattering, that is due to final--state interactions. However, as the meson pair production time in Fig.~\ref{npip} is practically instantaneous ($\sim 1/\sqrt{\hat{s}}$), while a much longer time ($\sim \sqrt{\hat{s}}/M^2$) is needed for the formation of a Reggeon by the secondary meson, there is insufficient time for a Reggeon emitted by one meson to interact with the other. More formally, this can be understood from the fact that, as shown by Mandelstam \cite{Mandelstam:1963cw}, the leading $s$ contribution in the case of additional Reggeon exchange comes from non--planar diagrams and not from planar graphs, of the type discussed in~\cite{Amati:1962nv}, and to which such a final--state meson--meson interaction corresponds. We refer the reader to~\cite{HarlandLang:2012qz} for a more detailed discussion of this.

However, following~\cite{HarlandLang:2011qd,HarlandLang:2012qz}, it may be necessary to introduce an additional suppression factor of the form $\sim \exp(-n(\hat{s}))$ to the cross section, corresponding to the small Poisson probability not to emit other secondaries in the $I\!\!P I\!\!P\to M_3M_4$ process at the initial meson pair production stage (rather than being due to final--state interactions between the mesons). Here $n(\hat{s})$ is the mean number of secondaries. We expect this to grow with the Pomeron-Pomeron energy, $\hat{s}=(P_1+P_2)^2$, as $n\simeq c\cdot\ln(\hat s/s_0)$,  with the  
coefficient $c\sim 0.5 -1$, see~\cite{HarlandLang:2011qd}. This factor may be described as the Reggeization of the meson $M^*$ exchange, which means that we now deal with non--local meson--Pomeron vertices and the $t$--channel meson $M^*$ becomes a non--local object, i.e. it has its own size. More precisely for the case of $\pi\pi$ CEP we can take 
\begin{align}\nonumber
n(\hat{s}) &=0 \qquad  &\sqrt{\hat{s}}<M_{f_2(1270)}\;,\\ \label{spipi}
n(\hat{s}) &=c\ln \left(\frac{\hat{s}}{s_0}\right) \qquad  &\sqrt{\hat{s}} \geq M_{f_2(1270)}\;,
\end{align}
with $s_0=M_{f_2(1270)}^2$. We take $c=0.7$ as a default value for definiteness, but we note that different choices are certainly possible. In this way, we account for the fact that we expect no additional suppression in the lower mass resonance region, where additional interactions at the meson pair production stage lead mainly to the
formation of resonances and not to the production of new secondaries.

A similar although slightly modified procedure is taken for the case of $K^+K^-$ CEP. In particular, we replace $M_{f_2(1270)}$ in (\ref{spipi}) with $M_{f_2(1525)}$, and account for the fact that we should expect the number of secondaries $n(s_{KK})$ to be a function of the free energy, $E_{\rm free}=M_{KK}-2M_K$, available for the creation of secondary particles, which can be numerically important because of the larger kaon mass $M_K$. In both cases the parameter $c$ in (\ref{spipi}) defines the strength of this additional `Poisson suppression', and can in principle be extracted from data, i.e. by measurements of the meson pair invariant mass distribution, although this is also sensitive to the form of the off--shell meson form factor described above, as we will see in Section~\ref{numer}.

\subsection{Reggeization of the exchanged meson}\label{secreg}

In principle, as $\hat{s}$ increases, we may have to account for the fact that the exchanged object in the $t$--channel is not a simple meson but can correspond to a whole family of exchanges; that is, the Reggeization of the intermediate meson. In such a case we may replace the meson propagator by\footnote{Sometimes in the literature a different form for the $\hat{t}$--dependence of the Reggeized meson exchange is used, see for example Eq (17) of~\cite{Yu:2011zu}, however our formulation is equivalent to this up to the (unknown) meson form factor $F_M(\hat{t})$, and so amounts to a simple redefinition of this object.}
\begin{equation}\label{exregge}
 \frac{1}{\hat{t}-M^2} \to  \frac{1}{\hat{t}-M^2}\left(\frac{s}{s_0}\right)^{\alpha_M(\hat{t})}\;,
\end{equation}
where $\alpha_M(\hat{t})$ is the Regge trajectory to which the exchanged meson belongs. However, some care is needed here, as it is not sufficient that the meson pair invariant mass $\sqrt{\hat{s}}$ should simply be large enough for such a description to be valid: rather, Reggeization occurs in the strongly ordered regime $\hat{s}\gg |\hat{t}|$, and so if we have $|\hat{t}| \sim \hat{s}$ at large $\hat{s}$ we cannot expect the replacement of (\ref{exregge}) to be justified. Indeed, for the experimentally relevant regime where the mesons are required to be produced quite centrally we may expect to be dominantly in this $|\hat{t}| \sim \hat{s}$ regime, with meson Reggeization becoming more important as the separation in rapidity between the mesons increases. To account for this, we may instead make the replacement
\begin{equation}\label{rasym}
 \frac{1}{\hat{t}-M^2} \to  \frac{1}{\hat{t}-M^2}\exp(\alpha_M(\hat{t})|y_{M_3}-y_{M_4}|)\;,
\end{equation}
where $y_{M_{3,4}}$ are the rapidities of the produced mesons. This has the correct Regge asymptotics in the $|\hat{t}|\ll \hat{s}$ regime, while giving no correction as the rapidity separation between the mesons tends to zero (i.e. for $|\hat{t}| \sim \hat{s}$), as it must\footnote{We note that (\ref{rasym}) can also be interpreted as the resummation of leading $\ln(1/x)$ contributions which may be enhanced with increased rapidity separation of the mesons.}. However, some care must still be taken, as the standard linear parameterization of the trajectory $\alpha_M(\hat{t})=\alpha_M(0)+\alpha'_M t$ can only be trusted for lower values of $|\hat{t}| \lesssim 1 \, {\rm GeV}^2$. Here, we choose to simply freeze the trajectory $\alpha_M(\hat{t})$ for $|t|>1\, {\rm GeV}^2$, but clearly there is a significant uncertainty in how to correctly include such an effect in the $|\hat{t}| \gtrsim 1 {\rm GeV}^2 $ region. More generally, it is not clear that this effect of meson Reggeization will be important in the 
relevant kinematic regime, when the mesons are produced relatively centrally, without a large separation in rapidity between them: as we will describe later, we allow the option to include this effect, in the way described above, in the new \texttt{Dime} MC, but by default it is not included.

\subsection{Secondary Reggeons}\label{secregs}

As well as the case of Pomeron exchange shown in Fig.~\ref{npip}, we must in general allow for the possibility of secondary Reggeon exchange between the protons and the produced mesons. This contribution will be subleading for sufficiently high c.m.s. energy squared $s$, but will not necessarily be completely negligible for the experimentally relevant region, in particular because in (\ref{namp}) the relevant subenergies $\hat{s}_{13},\, \hat{s}_{24}\sim M_X \sqrt{s}$ (and not $s$). Moreover, we will present a comparison in Section~\ref{numer} to ISR data, at $\sqrt{s}=62$ GeV, where it is crucial to include such secondary Reggeons.

Considering the case of $\pi^+\pi^-$ production, to achieve this in (\ref{namp}) we must make the replacement
\begin{equation}\label{reg}
i\sigma_0\bigg(\frac{\hat{s}_{13}}{s_0}\bigg)^{\alpha_{I\!\!P}(p^2_{1\perp})} \to \eta_{I\!\!P}\sigma_0^{I\!\! P}\bigg(\frac{\hat{s}_{13}}{s_0}\bigg)^{\alpha_{I\!\!P}(p^2_{1\perp})} + (\eta_f\,\sigma_0^f \pm \eta_\rho\,\sigma_0^\rho)\bigg(\frac{\hat{s}_{13}}{s_0}\bigg)^{\alpha_{I\!\!R}(p^2_{1\perp})} 
\end{equation}
where the `$\pm$' corresponds to the case that particle 3 is a $\pi^{\pm}$, with a similar replacement made for the $(ij)=(14)$ interaction.  The $\eta_i$ are the signature factors of the $I\!\!P$, $f_2$ and $\rho$ trajectories, given by $\eta_{I\!\!P}\approx i$, $\eta_\rho=-i-\tan (\pi\alpha_{I\!\!R}(0)/2)$ and  $\eta_f=i-\cot (\pi\alpha_{I\!\!R}(0)/2)$ while the normalisation factors $\sigma_0^i$ can be extracted from the fit of~\cite{Donnachie:1992ny} to the $\pi^\pm p$ cross sections. This gives $\sigma_0^f=31.79$ mb and $\sigma_0^\rho=4.23$ mb, while $\sigma_0^{I\!\! P}=$13.63 mb. For $K^+K^-$ production we have $\sigma_0^{f_2+a_2}=17.255$ mb, $\sigma_0^{\rho+\omega}=9.105$ mb and $\sigma_0^{I\!\! P}=$11.82 mb. Finally, the leading Reggeon trajectory is given by~\cite{Donnachie:1992ny}X
\begin{equation}
\alpha_{I\!\!R}(t)=0.55+0.93 \,t\;,
\end{equation}
where $t$ is measured in ${\rm GeV}^2$. We have also made the simplifying assumption that the form factor $F_p(t)$ for the Reggeon--proton coupling is the same as in the Pomeron case.

\subsection{Soft survival effects}\label{secsurv}
As discussed in Section~\ref{add}, we must account in our cross section calculation for the possibility that the colliding protons can interact in addition to the meson pair production process, spoiling the exclusivity of the final state. The probability that there is no additional proton--proton rescattering\footnote{We note that the discussion below is identical for the case of anti--protons.} is known as the `eikonal' survival factor $S^2_{\rm eik}$, see for example~\cite{Khoze:2002nf,Ryskin:2009tj,Ryskin:2011qe,Khoze:2014aca} for more details. As discussed in~\cite{HarlandLang:2010ep}, this factor is not a simple multiplicative constant, but rather depends on the distribution in impact parameter space of the colliding protons. In particular, in the simplest `one--channel' model, see e.g.~\cite{Khoze:2000cy} , which ignores any internal structure of the proton, we can write the average suppression factor as
\begin{equation}\label{S2}
\langle S^2_{\rm eik} \rangle=\frac{\int {\rm d}^2 {\bf b}_{1t}\,{\rm d}^2 {\bf b}_{2t}\, |\mathcal{M}(s,{\bf b}_{1t},{\bf b}_{2t})|^2\,{\rm exp}(-\Omega(s,b_t))}{\int {\rm d}^2\, {\bf b}_{1t}{\rm d}^2 {\bf b}
_{2t}\, |\mathcal{M}(s,{\bf b}_{1t},{\bf b}_{2t})|^2}\;,
\end{equation}
where ${\bf b}_{it}$ is the impact parameter vector of proton $i$, so that ${\bf b}_t={\bf b}_{1t}+{\bf b}_{2t}$ corresponds to the transverse separation between the colliding protons, with $b_t = |{\bf b}_t|$.  $\mathcal{M}(s,{\bf b}_{1t},{\bf b}_{2t})$ is the CEP amplitude (\ref{namp}) in impact parameter space, and $\Omega(s,b_t)$ is the proton opacity. Physically, ${\rm exp}(-\Omega(s,b_t))$ represents the probability that no inelastic scattering occurs at impact parameter $b_t$. 

While the rescattering probability only depends on the magnitude of the proton transverse separation $b_t$, the hard matrix element may have a more general dependence. More specifically, $\mathcal{M}(s,{\bf b}_{1t},{\bf b}_{2t})$ is the Fourier conjugate of the CEP amplitude (\ref{namp}), i.e. we have
\begin{equation}\label{Mfor}
\mathcal{M}(s,{\bf p}_{1_\perp},{\bf p}_{2_\perp})=\int {\rm d}^2{\bf b}_{1t}\,{\rm d}^2{\bf b}_{2t}\,e^{i{\bf p}_{1_\perp}\cdot {\bf b}_{1t}}e^{-i{\bf p}_{2_\perp}\cdot {\bf b}_{2t}}\mathcal{M}(s,{\bf b}_{1t},{\bf b}_{2t})\;,
\end{equation}
where the minus sign in the ${\bf p}_{2_\perp}\cdot {\bf b}_{2t}$ exponent is due to the fact that the impact parameter ${\bf b}_t$ is the Fourier conjugate to the momentum transfer ${\bf q}={\bf p}_{1_\perp}-{\bf p}_{2_\perp}$. We can therefore see that (\ref{S2}) is dependent on the distribution in the transverse momenta ${\bf p}_{i_\perp}$ of the scattered protons, being the Fourier conjugates of the proton impact parameters, ${\bf b}_{it}$. This connection can be made clearer by working instead in transverse momentum space, where we should calculate the CEP amplitude including rescattering effects, $T^{\rm res}$, by integrating over the transverse momentum ${\bf k}_\perp$ carried round the Pomeron loop (represented by the grey oval labeled `$S_{\rm eik}^2$' in Fig.~\ref{npip}). The amplitude including rescattering corrections is given by
\begin{equation}\label{skt}
\mathcal{M}^{\rm res}(s,\mathbf{p}_{1_\perp},\mathbf{p}_{2_\perp}) = \frac{i}{s} \int\frac{{\rm d}^2 \mathbf {k}_\perp}{8\pi^2} \;\mathcal{M}_{\rm el}(s,{\bf k}_\perp^2) \;\mathcal{M}(s,\mathbf{p'}_{1_\perp},\mathbf{p'}_{2_\perp})\;,
\end{equation}
where $\mathbf{p'}_{1_\perp}=({\bf p}_{1_\perp}-{\bf k}_\perp)$ and $\mathbf{p'}_{2_\perp}=({\bf p}_{2_\perp}+{\bf k}_\perp)$, while $\mathcal{M}^{\rm el}(s,{\bf k}_\perp^2)$ is the elastic $pp$ scattering amplitude in transverse momentum space, which is related to the proton opacity via
\begin{equation}\label{sTel}
\mathcal{M}_{\rm el}(s,t)=2s \int {\rm d}^2 {\bf b}_t \,e^{i{\bf q} \cdot {\bf b}_t} \,\mathcal{M}_{\rm el}(s,b_t)=2is \int {\rm d}^2 {\bf b}_t \,e^{i{\bf q} \cdot {\bf b}_t} \,\left(1-e^{-\Omega(s,b_t)/2}\right)\;,
\end{equation}
where $t=-{\bf k}_\perp^2$. We must add (\ref{skt}) to the `bare' amplitude excluding rescattering effects to give the full physical amplitude, which we can square to give the CEP cross section including eikonal survival effects
\begin{equation}\label{Tphys}
\frac{{\rm d}\sigma}{{\rm d}^2\mathbf{p}_{1_\perp} {\rm d}^2\mathbf{p}_{2_\perp}} \propto |\mathcal{M}(s,\mathbf{p}_{1_\perp},\mathbf{p}_{2_\perp})+\mathcal{M}^{\rm res}(s,\mathbf{p}_{1_\perp},\mathbf{p}_{2_\perp})|^2 \;,
\end{equation}
where here (and above) we have omitted the dependence of the cross section on all other kinematic variables for simplicity. In this way the expected soft suppression is given by 
\begin{equation}\label{seikav1}
\langle S_{\rm eik}^2\rangle= \frac{\int {\rm d}^2{\bf p}_{1_\perp}\,{\rm d}^2{\bf p}_{2_\perp}\,|\mathcal{M}(s,\mathbf{p}_{1_\perp},\mathbf{p}_{2_\perp})+\mathcal{M}^{\rm res}(s,\mathbf{p}_{1_\perp},\mathbf{p}_{2_\perp})|^2}{\int {\rm d}^2{\bf p}_{1_\perp}\,{\rm d}^2{\bf p}_{2_\perp}\,|\mathcal{M}(s,\mathbf{p}_{1_\perp},\mathbf{p}_{2_\perp})|^2}\;.
\end{equation}
It can readily be shown that (\ref{S2}) and (\ref{seikav1}) are equivalent. As we expect, the soft suppression factor depends on the proton transverse momenta, and so may have an important effect on the distribution of the outgoing proton ${\bf p}_{\perp i}$, via (\ref{Tphys}). A simplified approach, where the soft survival suppression is simply included in the CEP cross section as an overall constant factor will completely omit this effect. We make use of this formulation, in particular (\ref{Tphys}), in the \texttt{Dime} MC to give a full account of the survival factor and its effect on the distributions of outgoing proton momenta, which will be crucial in the presence of proton tagging. We will show in Section~\ref{numer} that the interference between the `screened' and `bare' amplitude in (\ref{Tphys}), given by (\ref{skt}) and (\ref{namp}), respectively, can in particular lead to some interesting diffractive dip phenomena, as was initially observed in~\cite{Khoze:2002nf}  (see also~\cite{HarlandLang:
2010ys,Harland-Lang:2013bya}).

Finally, we note that the formalism described above is only valid within the `one--channel' framework, which considers the pure elastic case, where the proton state is the correct degree of freedom for hadron--hadron scattering. More realistically, in particular to account for the possibility of (low mass) diffractive dissociation $p \to N^*$, a more sophisticated `multi--channel' framework is required, in which the incoming proton is considered to be in a coherent superposition of so--called diffractive eigenstates, which can each be described  by the above one--channel framework. The above formalism therefore still corresponds to the basic physics input into the model of soft diffraction that we use; the extension to the multi--channel case can be achieved in a quite straightforward manner, and is described in detail in~\cite{Khoze:2002nf,Ryskin:2009tj,Ryskin:2011qe}. We will make use of the two--channel model of~\cite{Khoze:2013dha} to calculate the eikonal survival factor. In this fit, the 
coupling of the Pomeron to the diffractive eigenstate $i=1,2$ is parameterized as
\begin{equation}\label{fi}
 F_i(t)=\exp(-(b_i(c_i-t))^{d_i}+(b_i c_i)^{d_i})\;,
\end{equation}
where the $b_i,c_i,d_i$ are extracted from data on hadronic scattering, and are given in~\cite{Khoze:2013dha}. To be consistent, we must then use this in our calculation of the `bare' CEP amplitude (\ref{namp}), i.e. for the proton form factor $F_p(t)$.

\section{\texttt{Dime MC}}\label{secdime}

We have implemented the model described in Section~\ref{theory} for exclusive meson pair production via double Pomeron/Reggeon exchange in the new \texttt{Dime} Monte Carlo, which is available via the \texttt{HepForge} webpage~\cite{dime}. The basic amplitude is given by (\ref{namp}) (with the replacement of (\ref{reg})), while the user can set the possible extensions and input parameters described in Sections~\ref{off}--\ref{secreg}. Specifically, the three different choices for the off--shell meson form factors (\ref{Fexp}--\ref{Fpow}) are available, and the parameters ($b_{\rm exp}$, $b_{\rm or}$, $a_{\rm or}$, $b_{\rm pow}$) may be set as input, with the default values given according to the comparison with ISR data described in Section~\ref{numer}. The Poisson suppression factor (\ref{spipi}) due to the requirement that no additional particles may be produced in the $I\!\! PI\!\! P \to M_3 M_4$ subprocess can be included or omitted, and the parameter $c$ can be set by the user, with the default given as 
described in Section~\ref{add}. 
Reggeization of the intermediate off--shell meson exchange, given by the approach described in Section~\ref{secreg}, is also available. Finally, soft survival effects are implemented as described in Section~\ref{secsurv}, including the full dependence on the ${\bf p}_{\perp i}$ of the outgoing protons, according to the new models described in~\cite{Khoze:2013dha}: a choice between any of the four models, as given by in Table 2 of~\cite{Khoze:2013dha}, is available to the user.

Currently, the \texttt{Dime} MC implements $\pi^+\pi^-$, $K^+K^-$, $\pi^0\pi^0$, $K^0K^0$ and $\rho_0(770)\rho_0(770)$ production. In the $\rho_0\rho_0$ case the mesons are decayed via $\rho_0 \to \pi^+\pi^-$, including the finite $\rho_0$ width, according to phase space only\footnote{A more complete treatment should account for the different $\rho$ polarization states, which may in general have distinct form factors $F_M(\hat{t})$, however given the lack of information about these we choose to ignore such possible polarization effects in the current version of the MC.}, while the factor $\sigma_0$ in (\ref{namp}) is set by default to the reasonable guess $\sigma_0^{I\!\! P}=10 \, {\rm mb}$, i.e. of order the $\pi^+\pi^-$ cross section, but taking a lower value due the larger $\rho_0$ mass. This somewhat arbitrary input is necessary due to the lack of $\rho_0 p$ scattering data with which to set the normalization (another reasonable choice may be to take $\sigma_0^{I\!\! P}=$13.63 mb as in $\pi^{\pm} p$ 
scattering~\cite{Donnachie:1992ny}). For 
$\rho_0\rho_0$ production, secondary Reggeons are not included and any spin effects are currently ignored in the production subprocess. Given the relative uncertainty in the $\rho_0\rho_0$ cross section normalisation, we also currently omit any effect from additional particle production, described in Section~\ref{add}, although this could in principle be included in the future.

\section{Numerical results}\label{numer} 

\subsection{Comparison to ISR data}
 
\begin{figure}
\begin{center}
\includegraphics[scale=0.65]{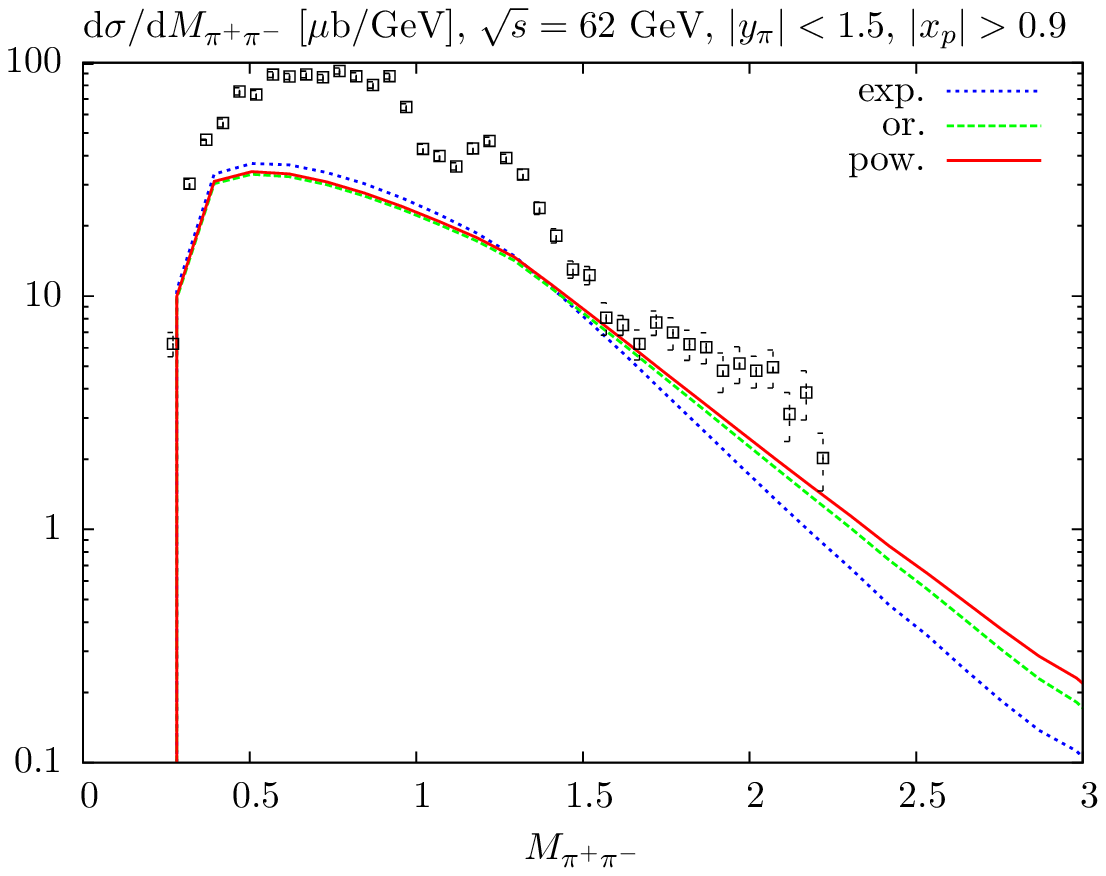}
\includegraphics[scale=0.65]{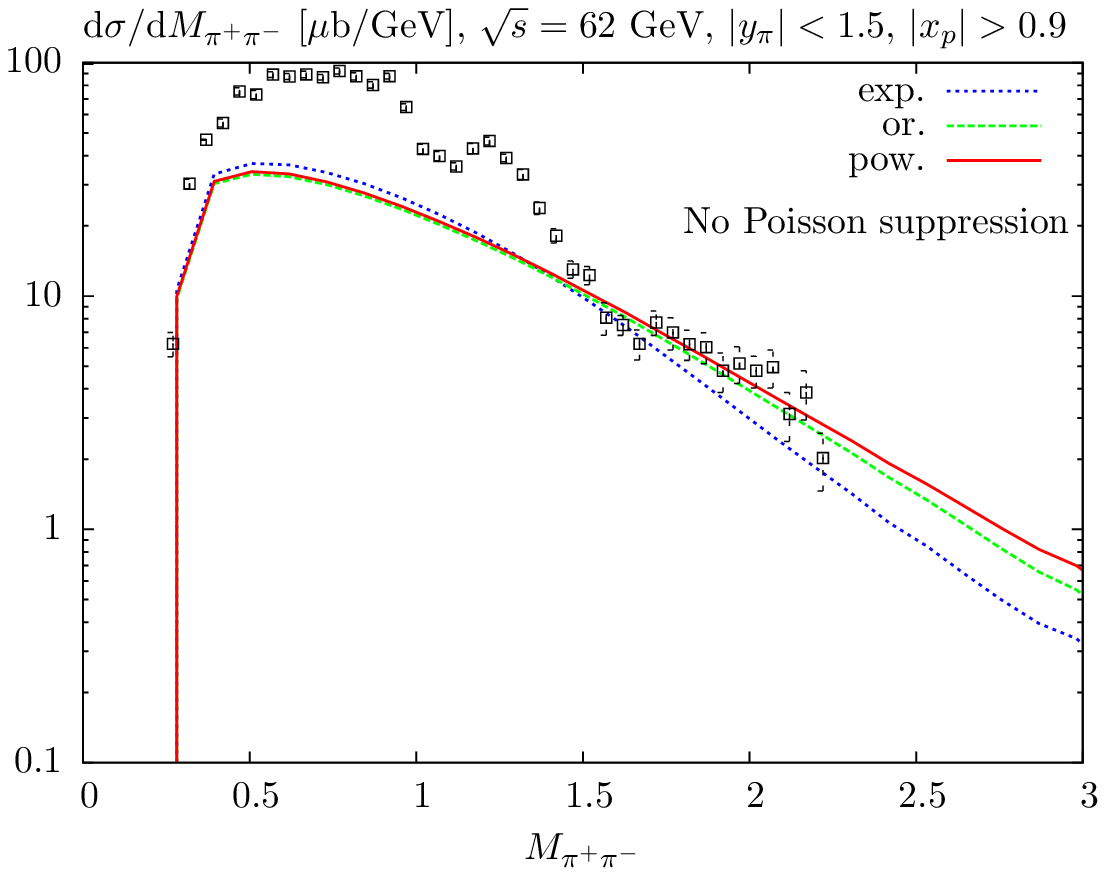}
\caption{Invariant mass distributions for $\pi^+\pi^-$ CEP at $\sqrt{s}=62$ GeV, compared to CERN--ISR data~\cite{Breakstone:1990at}. The theory curves are calculated as described in the text, using the three different parameterizations of the meson form factor $F_M(\hat{t})$ given in Section~\ref{off}. In the left/right figure the Poisson suppression described in Section~\ref{add} is included/excluded. In all cases, the pions are restricted to lie in the rapidity region $|y_{\pi}|<1.5$ and the cut $|x_p|>0.9$ is imposed on the outgoing protons.}\label{ISRpi}
\end{center}
\end{figure}

\begin{figure}
\begin{center}
\includegraphics[scale=0.65]{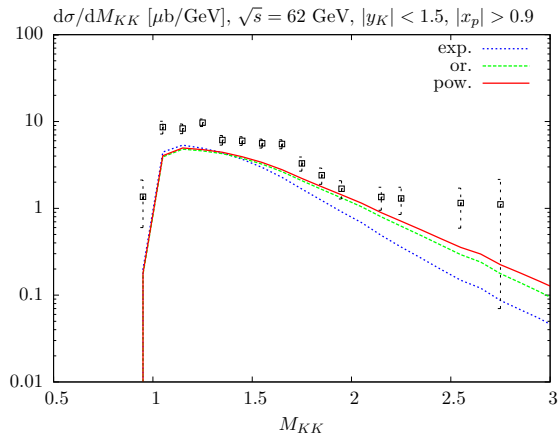}
\caption{Invariant mass distribution for $K^+K^-$ CEP at $\sqrt{s}=62$ GeV, compared to CERN--ISR data~\cite{Breakstone:1989ty}. The theory curves are calculated as described in the text, using the three different parameterizations of the meson form factor $F_M(\hat{t})$ given in Section~\ref{off}. The kaons are restricted to lie in the rapidity region $|y_{K}|<1.5$ and the cut $|x_p|>0.9$ is imposed on the outgoing protons.}\label{ISRK}
\end{center}
\end{figure}

In this section we compare the predictions\footnote{Unfortunately in all numerical results which follow, the bare cross sections were not normalized consistently with the two--channel model described in Section~\ref{secsurv}. All absolute cross section predictions presented in this paper should therefore be multiplied by a factor of 1.2, although the qualitative conclusions remain unchanged.} of this phenomenological model, described in the previous sections and implemented in the new \texttt{Dime} MC~\cite{dime}, with the existing low--energy ISR ($\sqrt{s}=62$ GeV) data on $\pi^+\pi^-$~\cite{Breakstone:1990at} and $K^+K^-$~\cite{Breakstone:1989ty} CEP. In Figs.~\ref{ISRpi} and~\ref{ISRK} we compare our predictions for the $\pi^+\pi^-$ and $K^+K^-$ invariant mass distributions with this ISR data, taking the three choices of meson form factor $F_M(\hat{t})$ described in Section~\ref{off}, calculated with model 1 of~\cite{Khoze:2013dha} for the eikonal survival factor, and with no meson Reggeization included. Fig.~\ref{ISRpi} (left/right) shows the prediction with the Poisson suppression (\
ref{spipi}) included/excluded, while in Fig.~\ref{ISRK} it is included. For a particular choice of parameters defined in (\ref{Fexp}--\ref{Fpow}), we can see that the non--resonant contribution to the data appears to be described reasonably well:  we take these values, 
given in Table~\ref{tpars}, as the default ones in \texttt{Dime MC}, and will use them in what follows, to present some representative predictions. However, it should be strongly stressed that a very precise comparison and extraction of these parameters is quite difficult from this somewhat limited data set. Firstly, most of the data lie in the resonant region: our model predictions only correspond to the non--resonant contribution to the  
experimental cross section, while for $\pi^+\pi^-$ production the $M_{\pi\pi}<2$ GeV region includes contributions from a number of resonances ($f_2(1270)$, $f_0(1370)$, $f_0(1500)$, $f_2'(1525)$, $f_2(1950)$,...) which overlap with each other, and similarly for the $K^+K^-$ case. It is not easy to disentangle the resonant and non--resonant contributions, and so our fit to the data can only be considered as a guideline. In particular, the $\pi^+\pi^-$ data are described to an acceptable degree out to the available $M_{\pi\pi} \approx 2$ GeV (where there may still be some resonant contribution) with and without the Poisson suppression (\ref{spipi}), with the same choice of parameters as in Table~\ref{tpars}. Secondly, we 
can see that the three different form factors give comparable fits to the data, with the possible exception of the exponential form factor when the Poisson suppression is included, which appears to undershoot the $\pi^+\pi^-$ data at higher $M_{\pi^+\pi^-}$. Moreover the ISR data, which only extend out to $M_{\pi^+\pi^-}\sim 2$ GeV (and similarly for the kaon case) are insensitive to the higher $k_\perp\gtrsim 1$ GeV behaviour of the form factors (\ref{Fexp}--\ref{Fpow}). Thus, the default values in Table~\ref{tpars} can only be taken as very rough guides, and the `true' values, and indeed general behaviour of these form factors may be quite different, in particular at higher $k_\perp$ (see the discussion in Section~\ref{off}).

\renewcommand{\arraystretch}{1.2}
\begin{table}
\begin{center}
\begin{tabular}{|c|c|c|c|}
\hline
$b_{\rm exp}$ $[{\rm GeV}^{-2}]$&$a_{\rm or}$ $[{\rm GeV}^{-1}]$&$b_{\rm or}$ $[{\rm GeV}^{-1}]$&$b_{\rm pow}$ $[{\rm GeV}^{2}]$\\
\hline
0.45 & 0.71 & 0.91 & 1.7\\
\hline
\end{tabular}
\caption{Parameters for different choices of meson off--shell form factor (\ref{Fexp}--\ref{Fpow}), extracted from ISR data~\cite{Breakstone:1990at,Breakstone:1989ty}.}\label{tpars}
\end{center}
\end{table}
\renewcommand{\arraystretch}{1}

\subsection{Limits from CDF diphoton data}\label{rCDF}
               
In~\cite{Aaltonen:2011hi} the observation of 43 $\gamma\gamma$ events with $|\eta(\gamma)|<1.0$ and  $E_T(\gamma)>2.5$ GeV, with no other particles detected in $-7.4<\eta<7.4$ was reported, corresponding to a cross section of $\sigma_{\gamma\gamma}= 2.48^{+0.40}_{-0.35} $ $({\rm stat})^{+0.40}_{-0.51}$ $ ({\rm syst})$ pb. In principle there could be some non--negligible contamination to this data from exclusive $\pi^0\pi^0\to 4\gamma$ production, with the photons in the $\pi^0$ decay merging or one photon being undetected, however this was determined experimentally to be very small, with $F_{\pi^0\pi^0}=N(\pi^0\pi^0)/N(\gamma\gamma)<0.34$, at 95\% C.L, and a best fit value of zero observed contamination ($F_{\pi^0\pi^0}=0$). Here `$N(\pi^0\pi^0)$' is the number of the observed $\pi^0\pi^0$ events, and similarly for $\gamma\gamma$. Once the difference in acceptance due to the $\pi^0\pi^0\to 4\gamma$ decay is accounted for, this may translate into a somewhat looser limit on the relative cross sections, 
although this effect should not be too dramatic, as to first approximation the photons from the decay of the very light pions will travel collinear to the original pion direction. This limit is in agreement with the perturbative results of~\cite{HarlandLang:2011qd}, which make the non--trivial prediction that the perturbative contribution to flavour--non--singlet meson pair production (such as $\pi^0\pi^0$) is dynamically suppressed, with in particular $\sigma(\pi^0\pi^0)/\sigma(\gamma\gamma)\sim 1\%$ within the CDF event selection.

However, we should also determine whether the `non--perturbative' approach described above would predict any significant $\pi^0\pi^0$ contamination in the kinematic region relevant to this CDF data. In Table~\ref{tcdf} we show the results of this model for the three different choices (\ref{Fexp}--\ref{Fpow}) of meson form factor described in Section~\ref{off}. We take model 1 of~\cite{Khoze:2013dha} for the eikonal survival factor, with no meson Reggeization included, and with the Poisson suppression (\ref{spipi}) included; if the Poisson suppression is excluded the resultant cross sections are larger by almost an order of magnitude, leading to even stronger constraints on the meson form factors. We can see immediately that the difference between these form factors is huge, spanning $\sim 4$ orders of magnitude. The reason for this is that these predictions are sensitive to relatively high $E_\perp>2.5$ GeV values, where the difference in the form factors is huge (recall also from (\ref{namp}) that the form 
factor enters to the fourth power in the cross section). Thus while the ISR data, see Figs.~\ref{ISRpi} and~\ref{ISRK}, are only sensitive to relatively low value of the meson $E_\perp$ (or equivalently, transverse momentum $k_\perp$) where the form factors are roughly similar in size, these high $E_\perp$ data provide a much more stringent constraint. In particular, we can see that both the power and Orear form factors, at least taking the default slope parameters of Table~\ref{tpars}, are completely inconsistent with the observed CDF limit on $\pi^0\pi^0$ CEP in this region. That is, ignoring acceptance effects for simplicity we have $\sigma(\pi^0\pi^0)<0.35\sigma(\gamma\gamma)\approx 0.8$ pb, and with a best fit value that is consistent with zero. Although as discussed above the true limit may be somewhat higher, this is nonetheless clearly much lower than the predicted $\pi^0\pi^0$ cross sections for the Orear and power form factors.

\begin{table}
\begin{center}
\begin{tabular}{|c|c|c|c|c|c|}
\hline
&CDF ($\gamma\gamma$)&Pert.&Exp.&Orear&Power\\
\hline
$\sigma$ [pb]&2.48&0.01&0.01&30&500 \\
\hline
\end{tabular}
\caption{Cross section predictions for $\pi^0\pi^0$ CEP, with $E_\perp(\pi^0)>2.5$ GeV, and $|\eta_\pi|<1$, calculated within the Regge--based approach discussed in this paper, for the three choices (\ref{Fexp}--\ref{Fpow}) of meson form factor described in Section~\ref{off}. Also shown are the predictions of the perturbative approach described in~\cite{HarlandLang:2011qd}, and, for the sake of comparison, the CDF measurement~\cite{Aaltonen:2011hi} of for $\gamma\gamma$ CEP within the same acceptance.}\label{tcdf}
\end{center}
\end{table}

Thus it appears that the CDF data tend to favour a `soft' exponential behaviour (\ref{Fexp}) for the meson factor, at least in this $E_\perp(\pi^0)>2.5$ GeV, $|\eta_\pi|<1$ region. However, as we shall comment below, new preliminary CDF data on $\pi^+\pi^-$ CEP at $\sqrt{s}$ = 900 and 1960 GeV, presented in~\cite{Albrow:2013mva,Mikeeds}, which extend out to $M_{\pi\pi} \sim 5$ GeV, and are therefore sensitive to a slightly lower region of meson $k_\perp$, appear in fact to favour the `Orear' behaviour of (\ref{Forexp}), and disfavour this exponential form factor. Clearly then this question of the behaviour of the coupling of the Pomeron to the off--shell meson, which we have parameterized in the simple forms given in Section~\ref{off}, is an uncertain one. More generally we should expect that, at sufficiently high meson $k_\perp$, the perturbative approach described in~\cite{HarlandLang:2011qd,Harland-Lang:2013ncy,Harland-Lang:2013qia} should be relevant. As we have discussed above, this approach predicts a 
strong dynamical suppression in the $\pi^0\pi^0$ CEP cross section that is not present at all in the non--perturbative model considered here. At sufficiently high meson $k_\perp$ this suppression should play a role, and the simplified behaviour given by the form factors of (\ref{Fexp}--\ref{Fpow}) should break down. It seems reasonable to assume that for the kinematic region probed by the CDF $\gamma\gamma$ data, for which we have $E_\perp(\gamma) >2.5$ GeV, a perturbative approach should be applied; indeed, the measured $\gamma\gamma$ cross section is in good agreement with the predictions of~\cite{HarlandLang:2010ep}, which apply exactly such an approach. Nonetheless, the question of when this transition to the `perturbative region' should occur, in particular in the case of $\pi^0\pi^0$ CEP (and other flavour--non--singlets), for which the pQCD--based approach predicts a strongly suppressed cross section, is an open one, and we may hope that future collider data on meson pair CEP will shed further light 
on this.

\subsection{Predictions for high energy colliders}

\begin{figure}
\begin{center}
\includegraphics[scale=0.65]{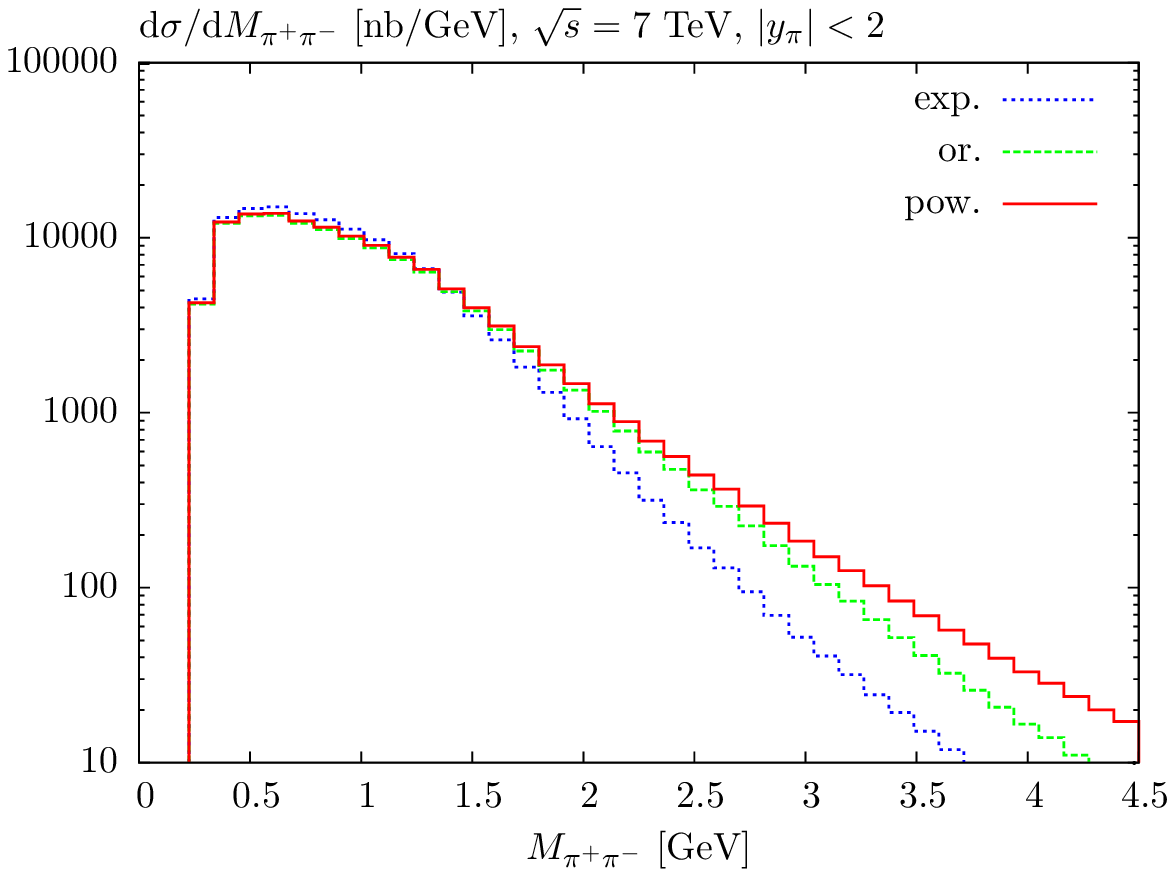}
\includegraphics[scale=0.65]{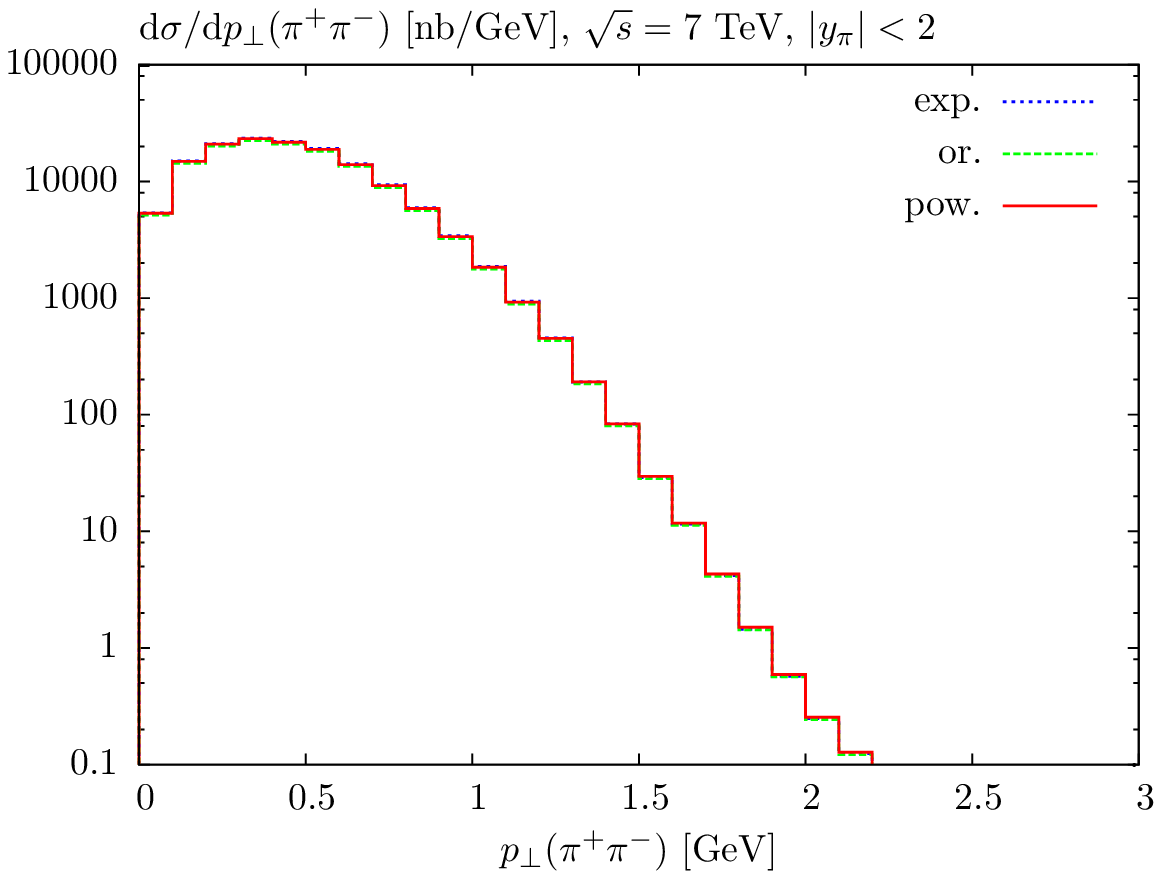}
\caption{Differential cross sections for $\pi^+\pi^-$ CEP at the LHC, $\sqrt{s}=7$ TeV, with respect to the $\pi^+\pi^-$ invariant mass and transverse momentum of the $\pi^+\pi^-$ system (note that in this case the three curves are almost identical). The pions are restricted to lie in the rapidity range $|y_\pi|<2$.}\label{LHC}
\end{center}
\end{figure}

\begin{figure}
\begin{center}
\includegraphics[scale=0.65]{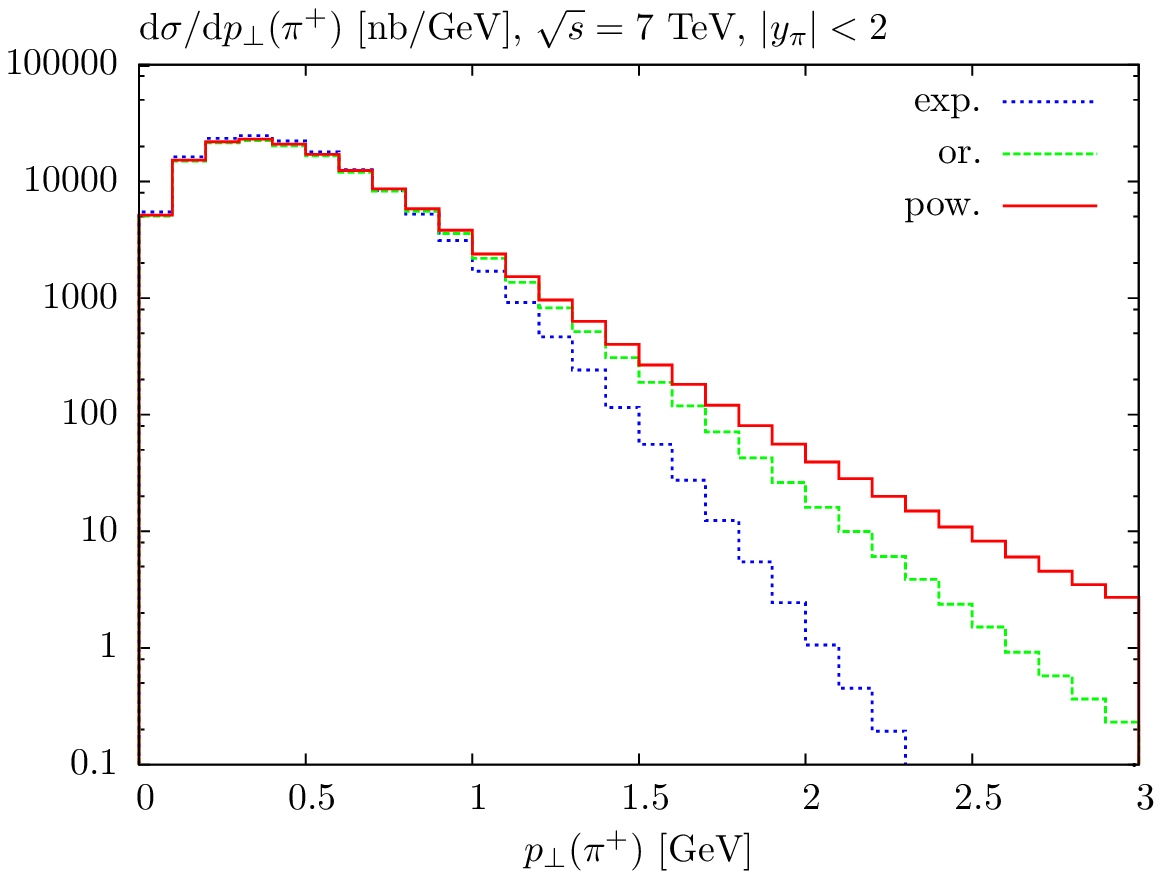}
\includegraphics[scale=0.65]{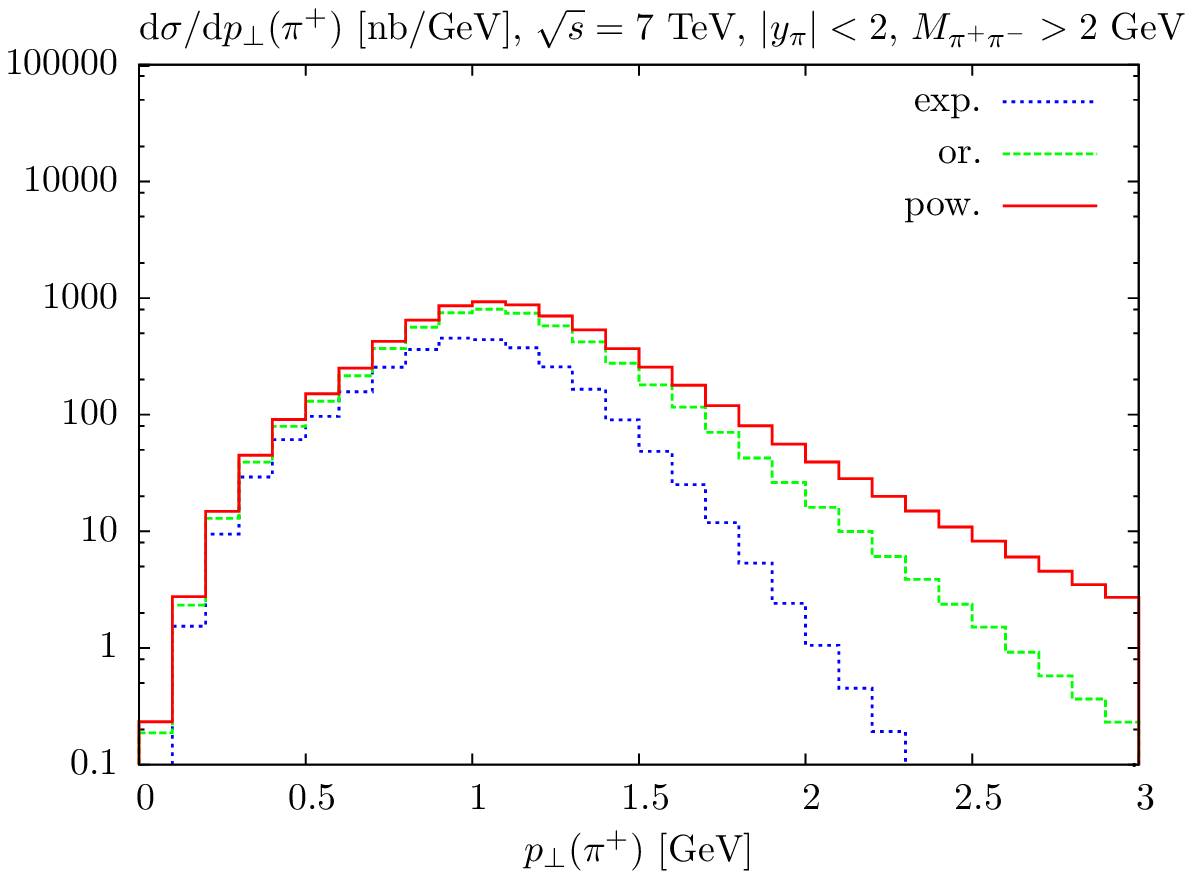}
\caption{Differential cross section for $\pi^+\pi^-$ CEP at the LHC, $\sqrt{s}=7$ TeV, with respect to the transverse momentum of the $\pi^+$. The pions are restricted to lie in the rapidity range $|y_\pi|<2$, and the additional cut of $M(\pi^+\pi^-)>2$ GeV is made in the right plot.}\label{LHC1}
\end{center}
\end{figure}

\begin{figure}
\begin{center}
\includegraphics[scale=0.65]{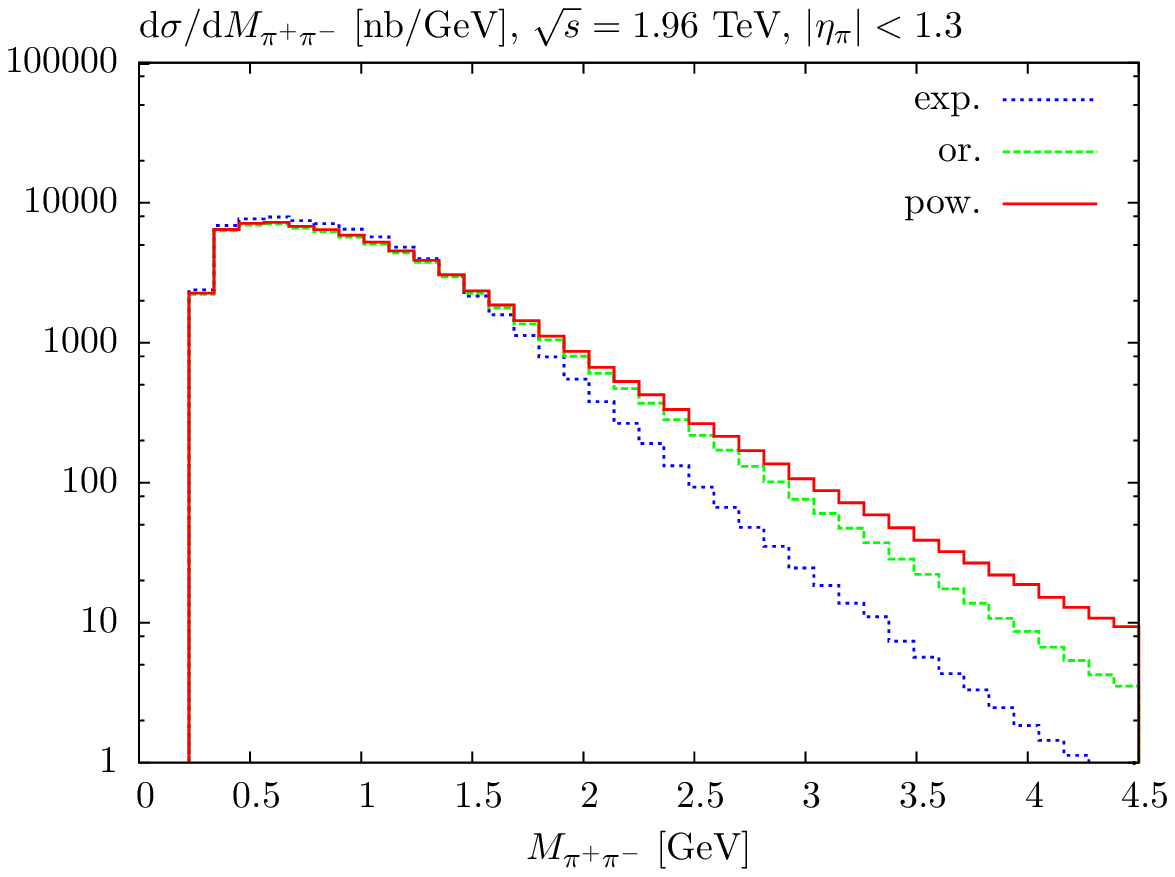}
\includegraphics[scale=0.65]{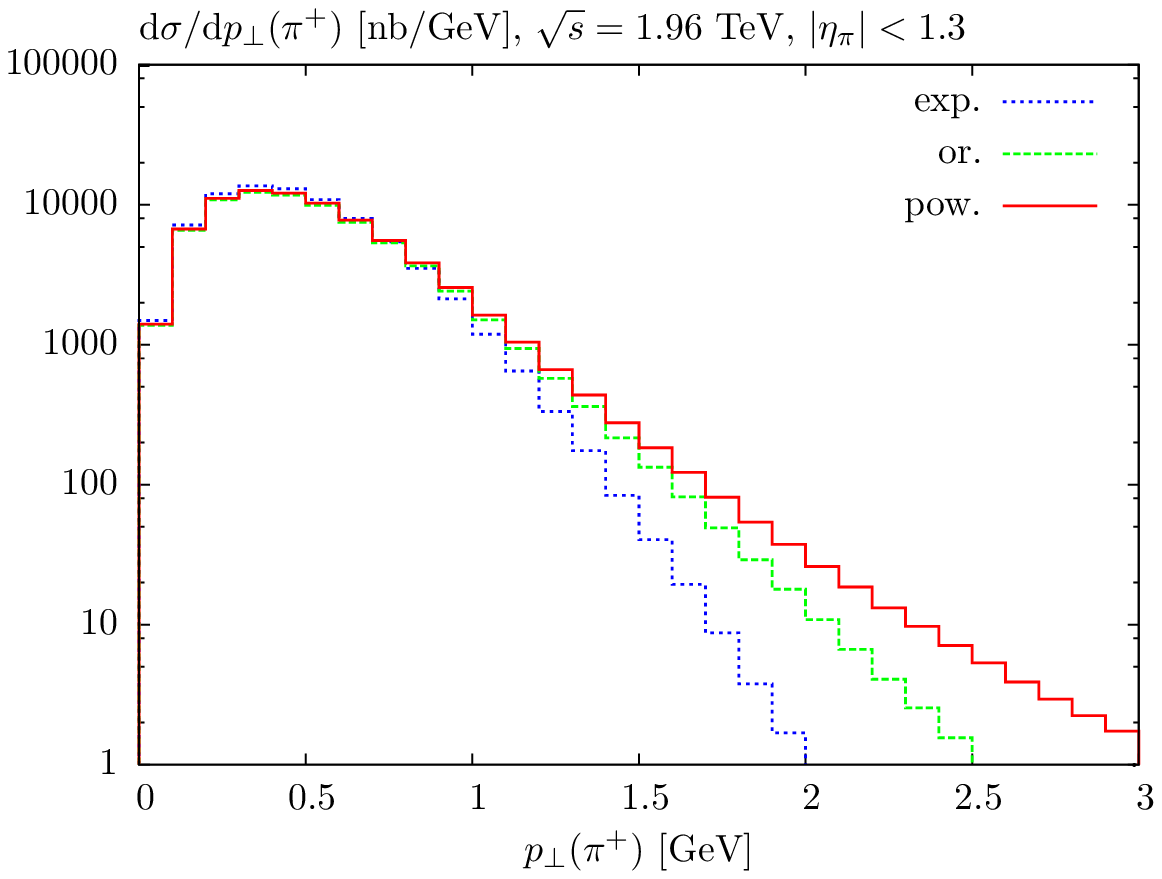}
\caption{Differential cross sections for $\pi^+\pi^-$ CEP at the Tevatron, $\sqrt{s}=1.96$ TeV, with respect to the $\pi^+\pi^-$ invariant mass and transverse momentum of the $\pi^+$. The pions are restricted to lie in the pseudorapidity range $|\eta_\pi|<1.3$.}\label{tev}
\end{center}
\end{figure}

In this section we present some selected numerical predictions for $\pi^+\pi^-$ and $K^+K^-$ production, for different c.m.s. energies and cuts on the final state particles, made throughout using the new \texttt{Dime} MC~\cite{dime}. We present predictions for the three choices of meson form factor described above, with the default parameters given in Table~\ref{tpars}: by comparing these predictions with future collider data, we may for example hope to shed some slight on the issues discussed in the previous section. We show in Figs.~\ref{LHC}--\ref{tev} cross section predictions for $\pi^+\pi^-$ CEP at the LHC ($\sqrt{s}=7$ TeV) and  Tevatron  ($\sqrt{s}=1.96$ TeV) collider energies, for the experimentally most relevant rapidity regions. All of the predictions which follow include the Poisson suppression of Section~\ref{add}, and do not include any meson Reggeization. Soft survival effects are calculated using model 1 of~\cite{Khoze:2013dha}, however we note that the distributions of the centrally produced 
particles which we will consider are highly insensitive to this choice (we will see later on that there is some dependence when the distribution of the outgoing protons is considered).  Predictions for other c.m.s. energies and model parameters can be readily made using the \texttt{Dime} MC: due to the energy dependence of the soft survival factor, which suppresses the cross section more with increasing $\sqrt{s}$, see e.g.~\cite{Khoze:2013dha}, we predict that the total meson pair CEP cross section will decrease gently with c.m.s. energy. This trend can be seen in Table~\ref{trts}, where we show the predicted $\pi^+\pi^-$, $K^+K^-$ and $\rho_0\rho_0$ cross sections at five different experimentally interesting $\sqrt{s}$ values, for the case that the pions/kaons are restricted to lie in the rapidity region $|\eta_{\pi,K}|<2.5$ (in the $\rho_0$ case this cut is imposed on the $\pi^+\pi^-$ decay products). These predictions are made using the `Orear' form factor (\ref{Forexp}); the $\pi^+\pi^-$ cross section, 
integrated down to zero transverse momentum, is largely insensitive to this choice, while there is some, $O(10\%)$ and $O(50\%)$, variation in the case of the higher mass $K^+K^-$ and $\rho_0\rho_0$ states, respectively. The predicted cross sections are very large, $O (\mu {\rm b})$, and so such processes represent very promising observables, even during low luminosity runs, with in particular the potential for making observations in the presence of tagged protons at CMS/ATLAS with the TOTEM/ALFA detectors being a very interesting possibility (see for example~\cite{Staszewski:2011bg,Sykoraeds}).

\renewcommand{\arraystretch}{1.2}
\begin{table}
\begin{center}
\begin{tabular}{|c|c|c|c|c|c|}
\hline
$\sqrt{s}$ [TeV]&0.5&0.9&1.96&7&14\\
\hline
$\sigma(\pi^+\pi^-)$ &28&23&20&17&16 \\
\hline
$\sigma(K^+K^-)$ &4.3&4.0&3.6&3.1&3.0 \\
\hline
$\sigma(\rho_0\rho_0)$ &0.29&0.28&0.26&0.26&0.25 \\
\hline
\end{tabular}
\caption{Cross sections (in $\mu$b) for $\pi^+\pi^-$, $K^+K^-$ and $\rho_0\rho_0$ production at different $\sqrt{s}$ values. The pions/kaons are restricted to lie in the rapidity region $|\eta_{\pi,K}|<2.5$, while this cut is imposed on the ($\pi^+\pi^-$) decay products of the $\rho_0$.}\label{trts}
\end{center}
\end{table}
\renewcommand{\arraystretch}{1}

\begin{figure}
\begin{center}
\includegraphics[scale=0.65]{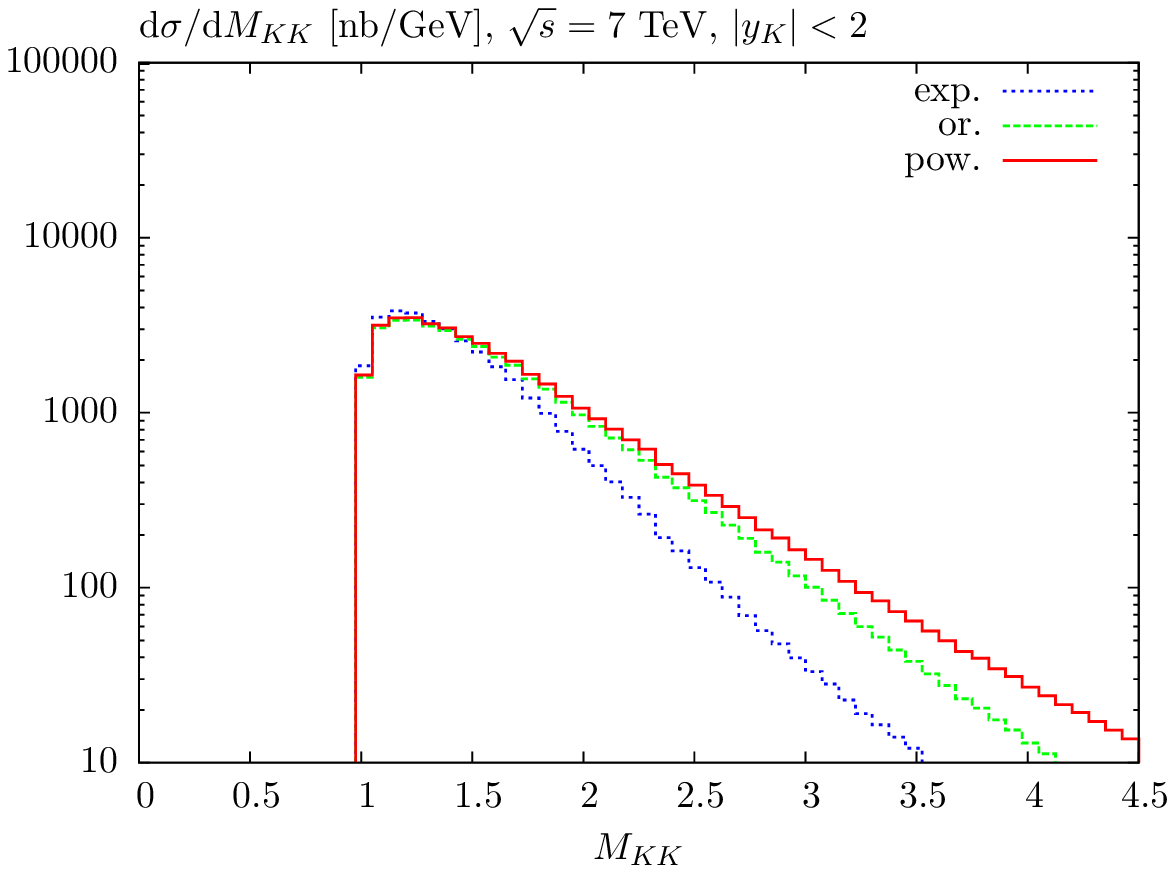}
\includegraphics[scale=0.65]{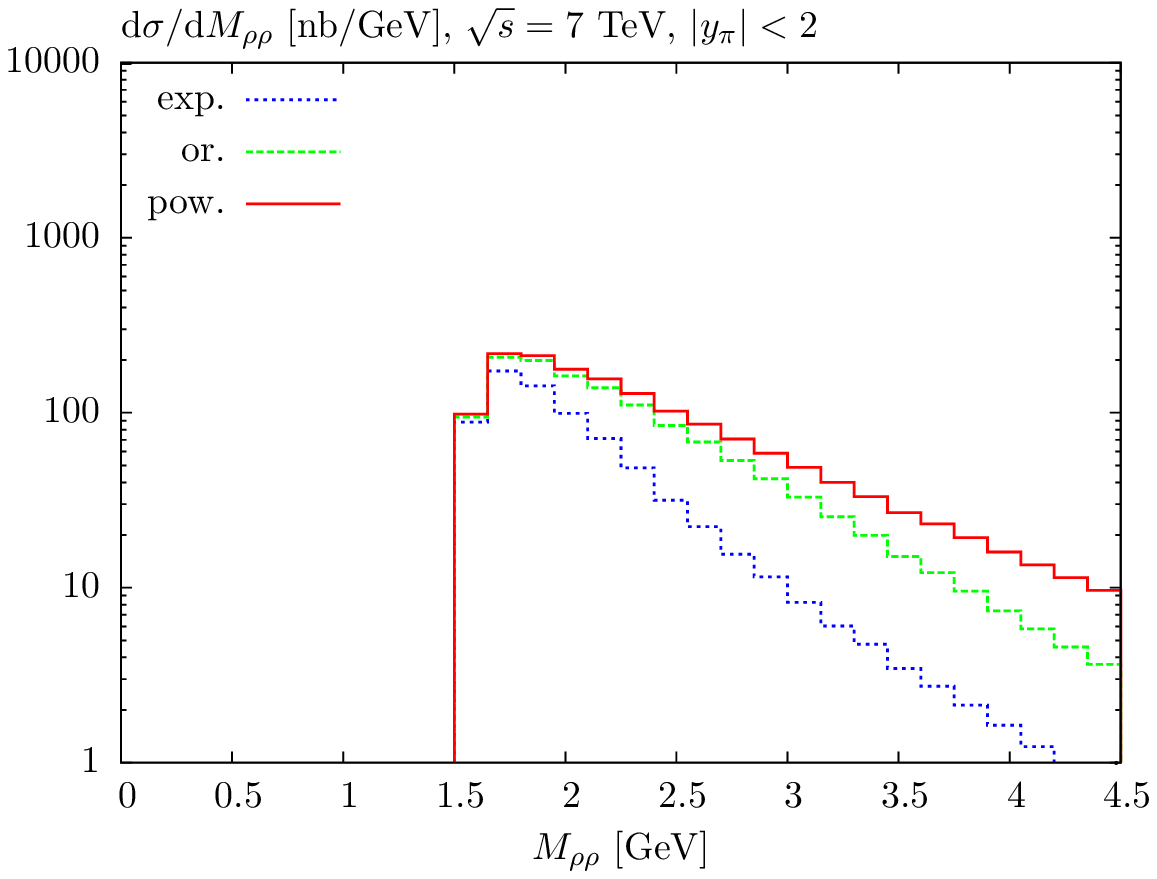}
\caption{Differential cross sections for $K^+K^-$ and $\rho_0\rho_0$ CEP at the LHC, $\sqrt{s}=7$ TeV, with respect to the meson pair invariant mass. The kaons and $\rho_0$ decay products ($\pi^+\pi^-$) are restricted to lie in the rapidity range $|y|<2$.}\label{LHCK}
\end{center}
\end{figure}

In Fig.~\ref{LHC} we show differential cross sections for $\pi^+\pi^-$ production at the LHC with respect to the invariant mass and transverse momentum of the $\pi^+\pi^-$ system. For this exclusive process, the $\pi^+\pi^-$ system recoils against the intact outgoing protons, and so this transverse momentum distribution is driven by the proton form factors $F_p(t)$ in (\ref{namp}). Thus the shape of the predicted differential cross section is essentially independent of the meson form factor, $F_M(\hat{t})$, taken in the production subprocess. Any significant deviation from these distributions for data selected without tagged protons (i.e. by rapidity vetoes within some acceptance regions), in particular the observation of a broader $p_\perp$ spectrum, may be evidence for a non--exclusive proton dissociative contribution to the data. In the case of the invariant mass distribution, there is a difference between the meson form factors, which becomes transparent above $M_{\pi^+\pi^-} \gtrsim 2$ GeV, to which the 
existing ISR data do not extend. This can also be seen in Fig.~\ref{LHC1}, for the distributions with respect to the transverse momentum $p_\perp$ of the $\pi^+$ (chosen for definiteness, although of course the $\pi^-$ distribution is completely equivalent), in particular in the region beyond $p_\perp(\pi^+)\approx 1$ GeV. We also show the case when an additional cut $M(\pi^+\pi^-)> 2$ GeV is imposed: this would in general be preferable in order to ensure we are safely away from the resonant region, and so isolate the non--resonant contribution. In Fig.~\ref{tev} we show similar distributions, but for the Tevatron ($\sqrt{s}=1.96$ TeV), while in Fig.~\ref{LHCK} we show results for $K^+K^-$ and $\rho_0\rho_0$ production at the LHC, and the conclusions are the same. 

\begin{figure}
\begin{center}
\includegraphics[scale=0.65]{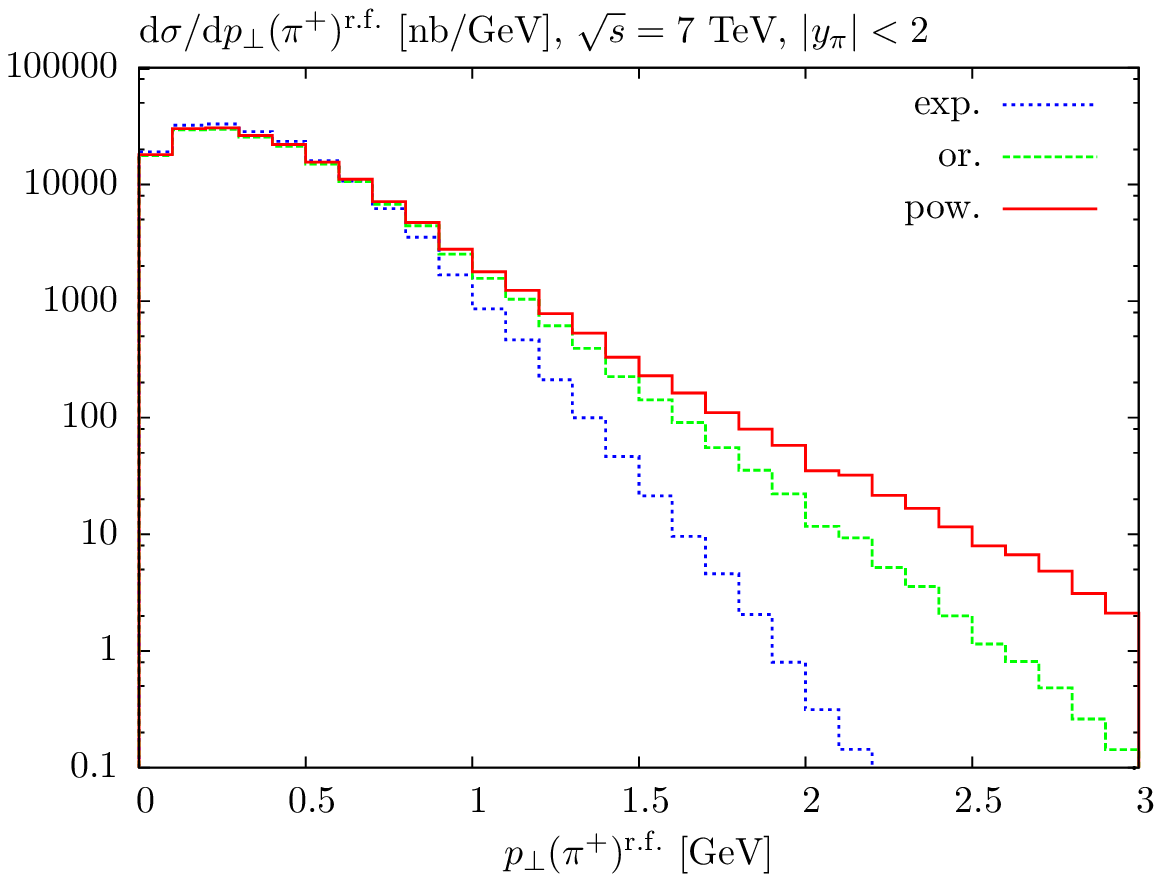}
\includegraphics[scale=0.65]{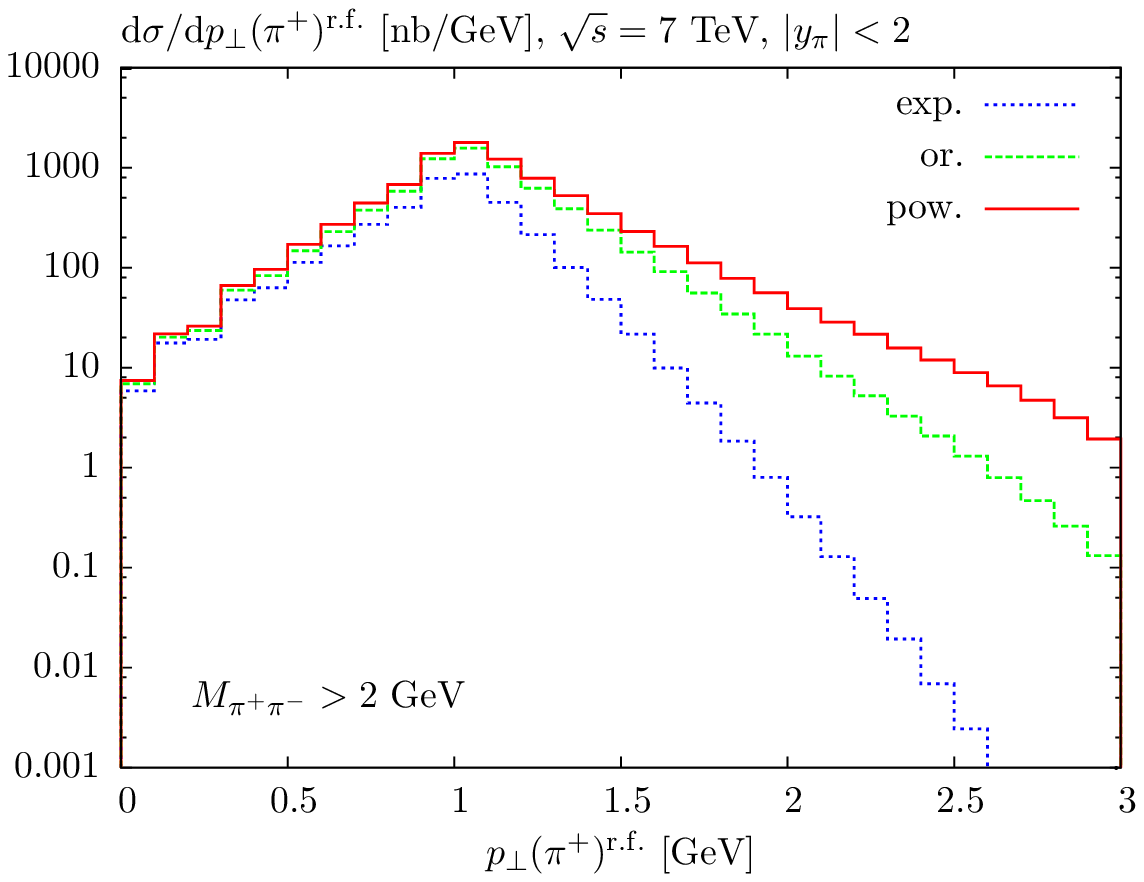}
\caption{Differential cross sections for $\pi^+\pi^-$ CEP at the LHC, $\sqrt{s}=7$ TeV, with respect to the transverse momentum of the $\pi^+$ in the $\pi^+\pi^-$ rest frame. The pions are restricted to lie in the rapidity range $|y_\pi|<2$.}\label{LHCr}
\end{center}
\end{figure}

The different form factor predictions can be further distinguished by considering the distribution of the meson ($\pi^+$, $K^+$...) transverse momentum in the meson pair rest frame, so that any contribution due to the non--zero $p_\perp$ of the meson pair (or equivalently, the outgoing protons) is subtracted. In Fig.~\ref{LHCr} we show predictions for this at the LHC, $\sqrt{s}=7$ TeV, for $\pi^+\pi^-$ production, and we find that, in particular for the exponential meson form factor, which has the softest predicted $p_\perp$ distribution, the difference between the form factor predictions is increased. More generally, it is preferable to consider such a variable as it is independent of the proton $p_\perp$ distribution (which is dependent on the proton form factors $F_p(t)$) and so it will only be driven by the physics of the meson pair production subprocess, i.e. the choice of meson form factor. 


\begin{figure}
\begin{center}
\includegraphics[scale=0.65]{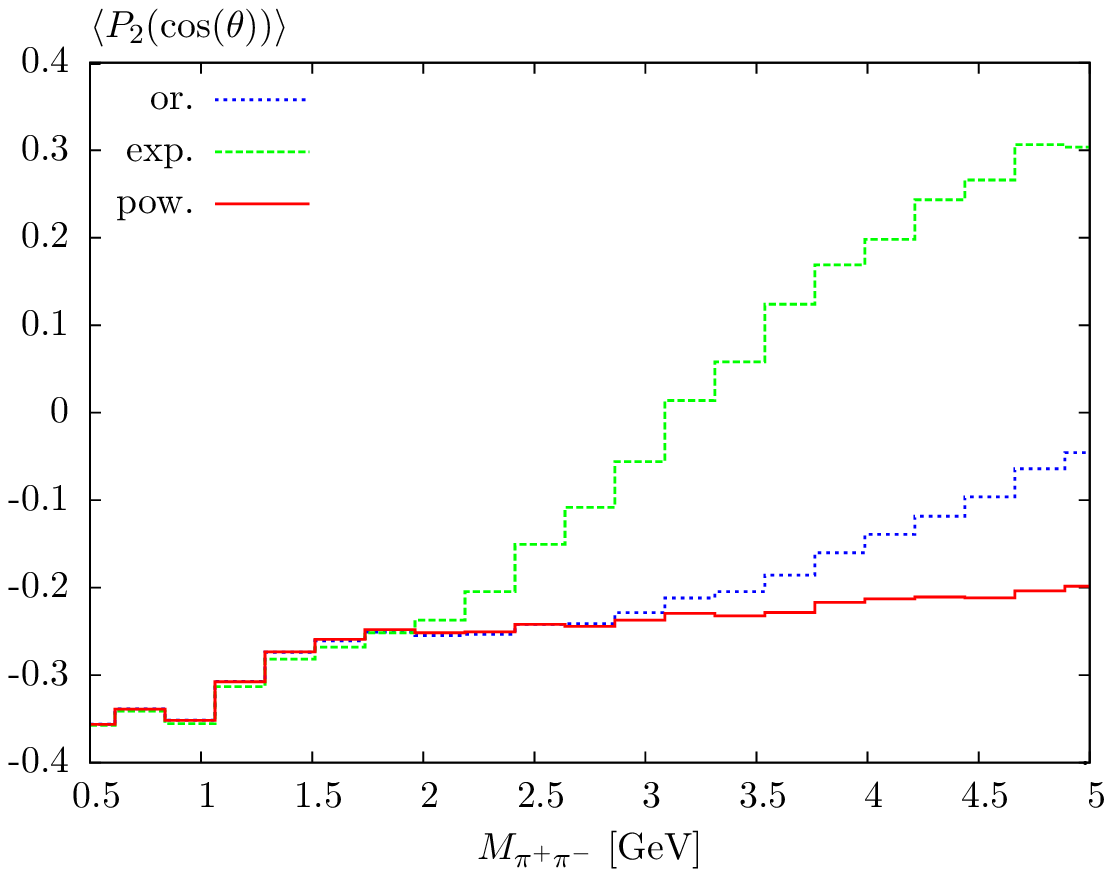}
\includegraphics[scale=0.65]{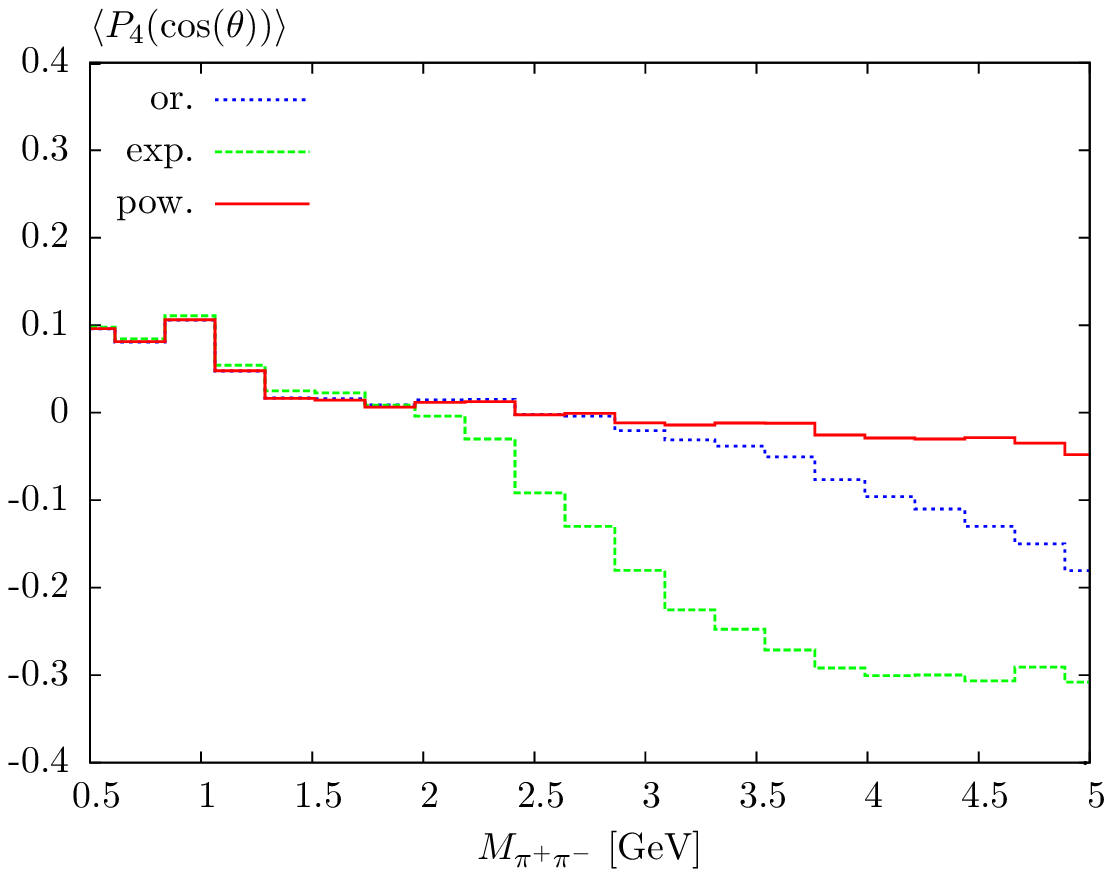}
\caption{Expectation values of the $l=2,4$ Legendre polynomials $P_l(\cos\theta)$, where $\theta$ is the angle of the $\pi^+$ with respect to the beam axis, in the $\pi^+\pi^-$ rest frame. The predictions shown correspond to $\sqrt{s}=1.96$ TeV and for the additional cuts $p_\perp(\pi)>0.4$ GeV, $|\eta(\pi)|<1$, $|y(\pi^+\pi^-)|>1.0$ imposed, as in~\cite{Albrow:2013mva,Mikeeds}.}\label{pls}
\end{center}
\end{figure}

Another observable which can be highly sensitive to the choice of meson form factor $F_M(\hat{t})$ is the angular distribution of the mesons in the pair rest frame. Higher values of the meson transverse momentum $p_\perp$ are disfavoured, and so larger values of the pair invariant mass will be preferentially produced by large rapidity separations between the mesons, where the corresponding meson $p_\perp$ is minimized. Such a 
configuration is equivalent to higher values of $|\cos\theta|$ in the meson pair rest frame, and so we will expect this behaviour to affect the meson angular distributions in a $M_X$ dependent way, with the precise quantitative prediction depending on the choice of  $F_M(\hat{t})$. A particularly transparent way to examine the mass dependence of these distributions is to consider the expectation values of the Legendre polynomials $P_l(\cos\theta)$, and so in Fig.~\ref{pls} we show this for $\pi^+\pi^-$ production at $\sqrt{s}=1.96$ TeV for the two lowest non--trivial $l=2,4$ cases\footnote{In Fig.~\ref{pls} the angle $\theta$ is defined with respect to the beam axis, in the $\pi^+\pi^-$ rest frame. It is also possible to consider a related observable, defined with respect to the incoming Pomeron/Reggeon exchange, however this cannot be determined experimentally in the absence of proton tagging.} (note that the predicted angular distributions are even in $\cos\theta$ and so the odd $l$ contributions vanish), 
using the three choices of pion factor given in Section~\ref{off}. The difference between the form factors is immediately clear, and thus such observables may prove very useful in distinguishing between these choices (we also find that these distributions are largely unaffected by the soft survival factor). We note that preliminary measurements of these distributions by the CDF collaboration have been presented in~\cite{Albrow:2013mva,Mikeeds}, and these are in quite encouraging agreement with the predictions of Fig.~\ref{pls} (which apply the same cuts on the pions as in the CDF analysis) for the case of the `Orear' (\ref{Forexp}) form factor\footnote{We note in passing that one interesting measurement reported in~\cite{Albrow:2013mva,Mikeeds} is a new 
limit on the $\chi_{c0}$ CEP cross section, via the $\chi_{c0} \to \pi^+\pi^-$, $K^+K^-$ channels, of ${\rm d}\sigma/{\rm d}y|_{y=0}(\chi_{c0}) \lesssim 20 $ nb at 90\% confidence. This appears to suggest a somewhat larger contribution from the higher spin $\chi_{c(1,2)}$ states to the $\chi_{cJ} \to J/\psi \gamma$ combined CDF cross section measurement of~\cite{Aaltonen:2009kg} than that predicted in~\cite{HarlandLang:2010ep}, and a similar trend is also seen in the preliminary LHCb data~\cite{LHCbconf}. To further clarify this issue we note that an observation of $\chi_{c0} \to \pi^+\pi^-$ may be possible if an additional constraint is imposed on the minimum pion $p_\perp$ (see also~\cite{HarlandLang:2012qz}) or, equivalently, maximum $|\cos \theta|$; while the form factor $F_M(\hat{t})$ will lead to a preference for larger $\cos \theta$ values as $M_{\pi^+\pi^-}$ is increased, the isotropic $\chi_{c0}$ decay distribution will not, and so by requiring $|\cos \theta|<0.6$ (say) the continuum background will 
be reduced preferentially, and an increased $S/\sqrt{B}$ may be achievable.} (there is good agreement as well for the higher $l=6,8$ terms, which are found to be small in the data). While this data appears to favour such a form factor, we recall from Section~\ref{rCDF} that such a form factor appears to be in strong conflict with the earlier CDF limit~\cite{Aaltonen:2011hi} on $\pi^0\pi^0$ CEP for $E_\perp(\pi^0)>2.5$ GeV and $|\eta_\pi|<1$.

\begin{table}
\begin{center}
\begin{tabular}{|c|c|c|c|c|c|}
\hline
$\sqrt{s}$ [TeV]&0.5&0.9&1.96&7&14\\
\hline
model 1 &15&13&10&7.6&6.4 \\
\hline
model 2 &23&21&17&13&11 \\
\hline
model 3 &15&14&12&10&8.8 \\
\hline
model 4 &15&14&12&9.1&7.9 \\
\hline
\end{tabular}
\caption{Soft suppression $\left\langle S^2_{\rm eik}\right\rangle$ (in \%), defined in (\ref{seikav1}), and calculated using the four soft models described in~\cite{Khoze:2013dha}, for $\pi^+\pi^-$ CEP at different c.m.s. energies. The pions are restricted to lie in the rapidity region $|y_\pi|<2.5$.}\label{tsurv}
\end{center}
\end{table}

Finally, we consider in more detail the effect of the soft survival factor, see section~\ref{secsurv}, on the cross section predictions. In Table~\ref{tsurv} we show predictions for the average suppression factors $\left\langle S^2_{\rm eik}\right\rangle$, defined in (\ref{seikav1})\footnote{More precisely, it is Eq. (11) of~\cite{Khoze:2013dha} which is used. We note that the averaged survival factors quoted here cannot be used directly to give the relative cross section predictions for the different models of~\cite{Khoze:2013dha}, as the `bare' cross section itself, i.e. the denominator of this Eq. (11), also depends on the model choice through the different couplings of the Pomeron to the Good--Walker eigenstates in each model.} by which soft survival effects will suppress the total $\pi^+\pi^-$ cross sections with $|y_\pi|<2.5$. We show these for four different soft models described in~\cite{Khoze:2013dha} (see Section~\ref{secsurv}). As discussed in~\cite{Khoze:2013dha}, these models corresponds to 
different input parameters and parameterizations of the Pomeron coupling to Good--Walker eigenstates, and these all provide a good description of the available soft hadronic data on elastic and diffractive scattering. On the other hand, we can see that the predictions for the survival factor vary by as much as a factor of $2$ between these model choices, demonstrating the uncertainty in the current models of soft physics.

\begin{figure}
\begin{center}
\includegraphics[scale=0.65]{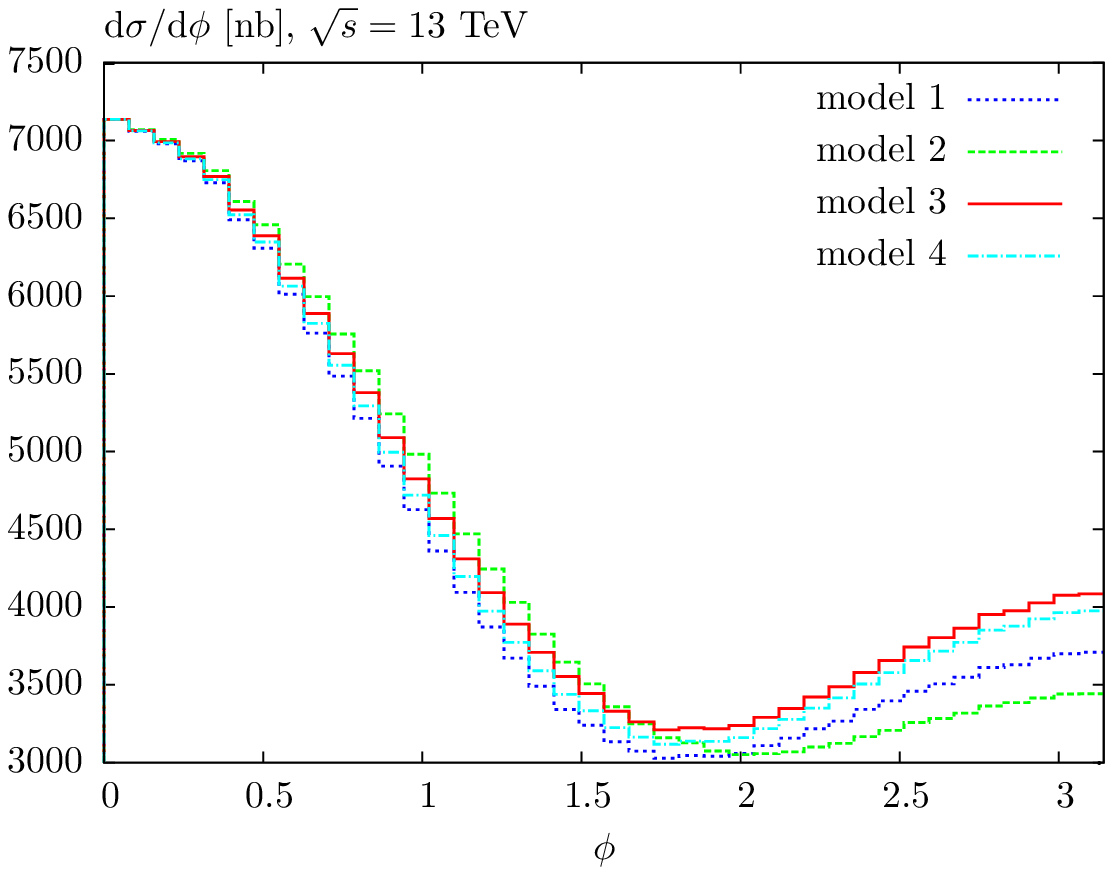}
\includegraphics[scale=0.65]{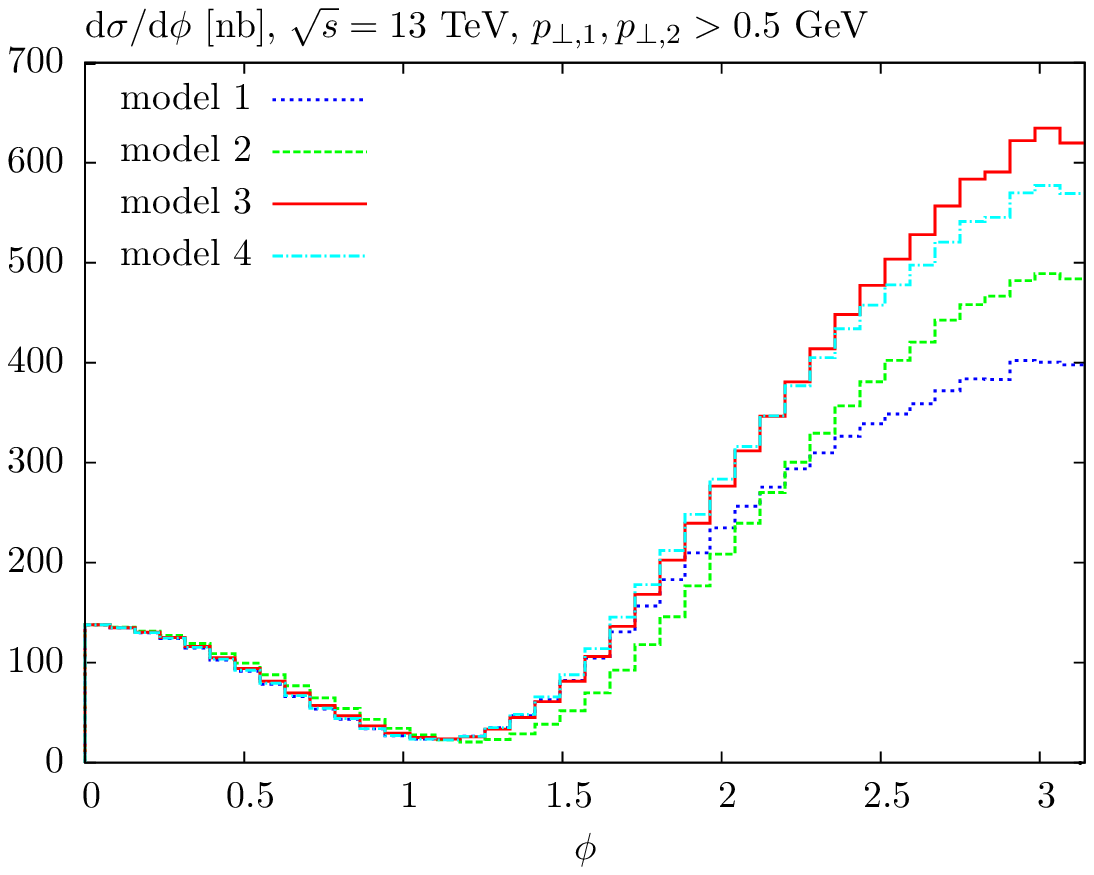}
\includegraphics[scale=0.65]{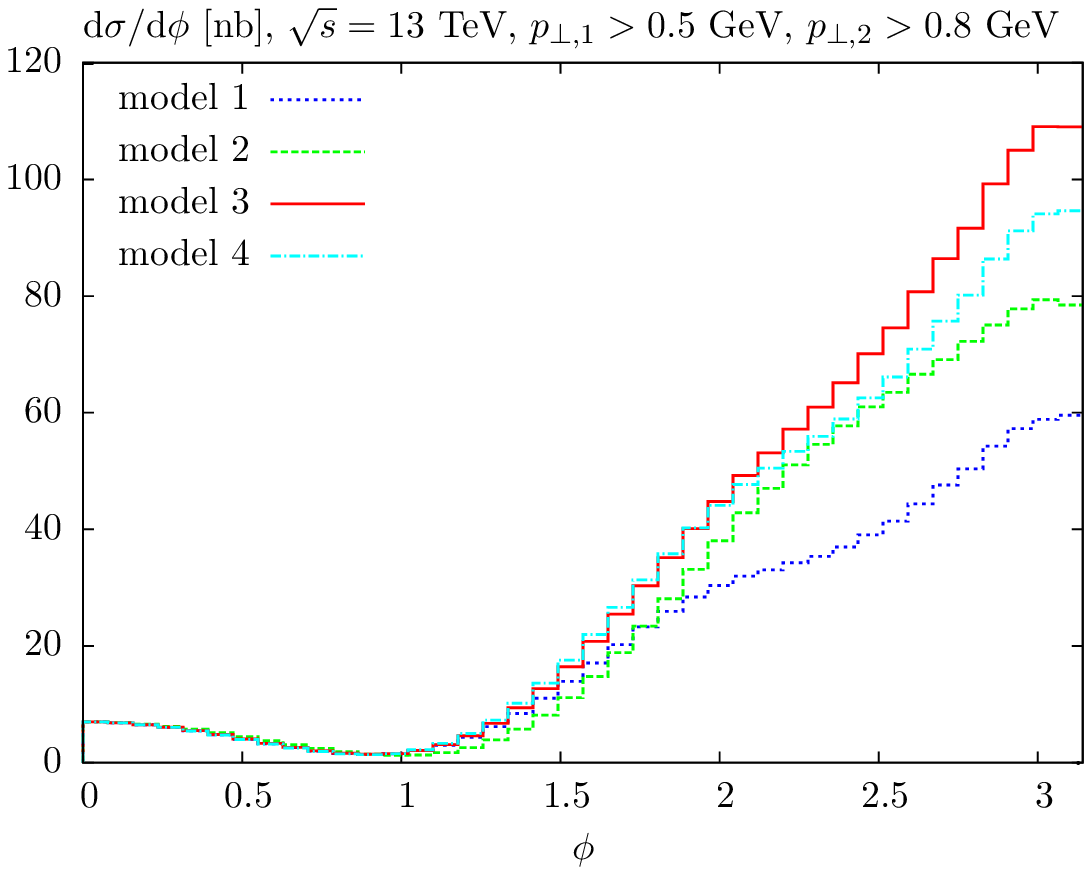}
\includegraphics[scale=0.65]{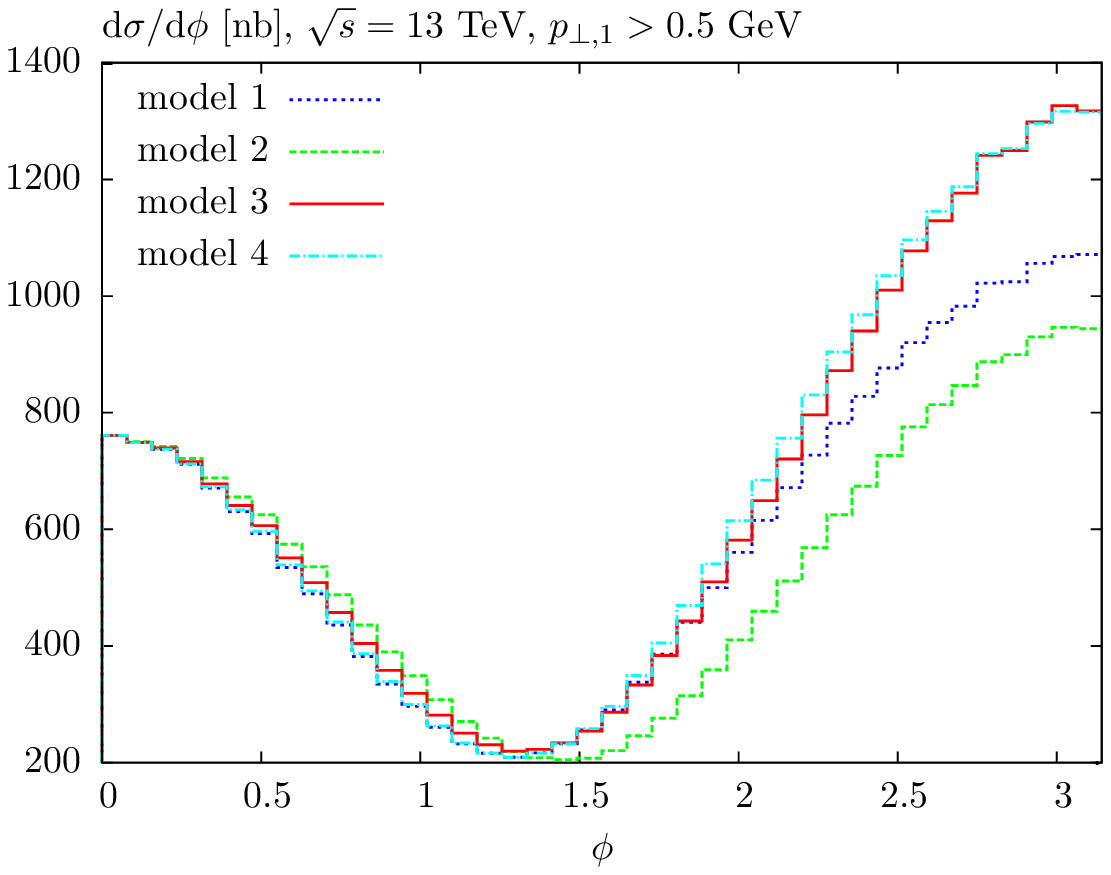}
\caption{Differential cross section ${\rm d}\sigma/{\rm d}\phi$, where $\phi$ is the azimuthal angle between the outgoing proton $p_\perp$ vectors, at the $\sqrt{s}=13$ TeV LHC, for the four soft models of~\cite{Khoze:2013dha}. Results are also shown for different cuts on the magnitude of the proton $p_\perp$, and for a cut $|y_\pi|<2$ on the centrally produced pions. For display purposes the predictions are normalized in the first $\phi$ bin, to the model 1 predictions.}\label{LHCphi}
\end{center}
\end{figure}

\begin{figure}
\begin{center}
\includegraphics[scale=0.65]{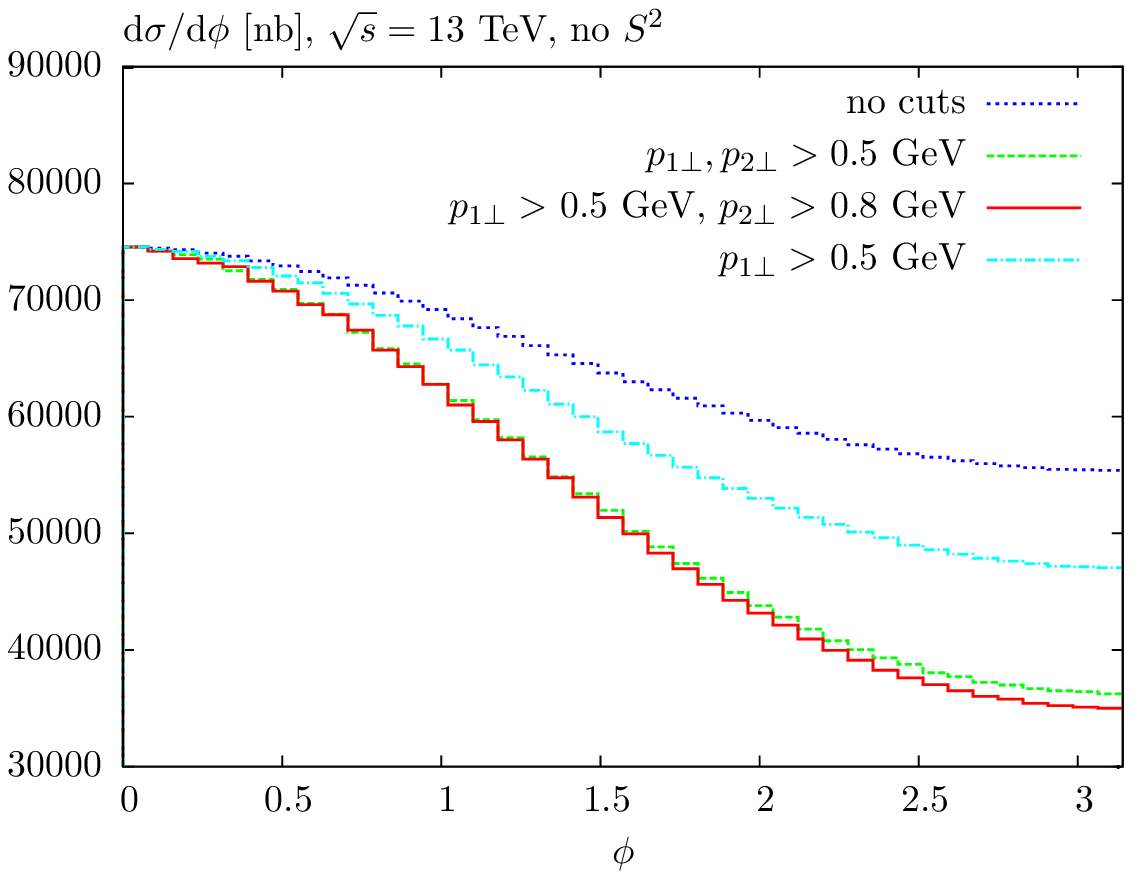}
\caption{Differential cross section ${\rm d}\sigma/{\rm d}\phi$, where $\phi$ is the azimuthal angle between the outgoing proton $p_\perp$ vectors, at the $\sqrt{s}=13$ TeV LHC, with soft survival effects omitted. Results are shown for the four choices of cuts shown in Fig.~\ref{LHCphi}, and for a cut $|y_\pi|<2$ on the centrally produced pions.  For display purposes the predictions are normalized in the first $\phi$ bin, to the prediction where no cuts are applied to the outgoing protons.}\label{LHCphins}
\end{center}
\end{figure}

We may also consider what additional information can be provided on the soft survival factor by measuring the momenta of the outgoing intact protons. It is in particular interesting to consider the distribution in the azimuthal angle between the $p_\perp$ vectors of the outgoing protons (see for instance~\cite{Khoze:2002nf,HarlandLang:2010ys,Harland-Lang:2013bya}) which is in general sensitive to both the structure of the production subprocess and spin/parity of the centrally produced state, as well as soft survival effects. In Fig.~\ref{LHCphi} we show this distribution at the LHC ($\sqrt{s}=13$ TeV) for the four different soft models described in~\cite{Khoze:2013dha}. While for the full cross section it appears that there is only a fairly small difference in shape between the different models, once cuts are placed on the magnitude of the proton $p_\perp$, this difference becomes more apparent. Moreover, we can observe a very distinct `diffractive' dip structure, with the distributions reaching a minimum at 
a particular value of $\phi$. This is a consequence of the destructive interference between the screened and bare amplitudes in (\ref{Tphys}), which becomes particularly pronounced at higher proton $p_\perp$, corresponding to a less peripheral interaction where survival effects are stronger. For a particular value of $\phi$ this interference is strongest, resulting in the observed minimum in the $\phi$ distribution (such an effect was predicted in~\cite{Khoze:2002nf}, see also~\cite{HarlandLang:2010ys,Harland-Lang:2013bya}). For the sake of comparison, in Fig.~\ref{LHCphins} we show the $\phi$ distributions for these different cuts, without survival effects included (i.e. simply taking the `bare' amplitude of (\ref{namp})), and we can see that this dip behaviour disapears completely. As the form of the screened amplitude depends on the particular soft model, we may expect the position and depth of this minimum to be sensitive to this, as well as depending on the particular cuts imposed on the proton $p_\perp$. In fact, it appears from Figs.~\ref{LHCphi} that the position of the minimum does not depend too strongly on the choice of model, but nonetheless the overall shape of the $\phi$ distribution does show some variation. We note that these distributions are largely independent of the details of the meson production subprocess (i.e. the shape taken for meson form factor, although for completeness we note that the exponential form factor (\ref{Fexp}) is taken here), and so represent a potentially unique handle with which to test the different available models for soft proton interactions. For this reason, the observation of, for example $\pi^+\pi^-$ and/or $K^+K^-$ CEP with tagged protons would provide valuable insight into this issue.

\section{The effect of proton dissociation}\label{secdiss}

As discussed in the introduction, a LRG veto on additional particles in a certain rapidity region is commonly used to experimentally select exclusive events. However, without tagging the outgoing intact protons it is impossible to guarantee that these  will be purely exclusive, and there will always in general be some contamination from events where either or both protons dissociate. It is therefore crucial to understand how important such a contribution will be when comparing any data selected with a LRG veto with the purely exclusive prediction presented here.

To evaluate the size of this possible contamination at the LHC, we can use the recent HERA measurements, by the H1 collaboration, of the elastic and proton--dissociative cross sections for the photoproduction of $J/\psi$ mesons~\cite{Alexa:2013xxa} as a guide. In this case, it is found that the ratio of proton-dissociative (pd) to elastic cross sections is quite large, with $\sigma_{\rm pd}/\sigma_{\rm el}\sim 0.9$, for a proton--dissociative system $Y$ of mass $M_Y<10$ GeV. If we now consider the LHC, the range of allowed $M_Y$ can be estimated by considering the case that the final--state particles in the dissociative system have some average transverse momenta $\langle k_\perp \rangle$, for which the system spans a rapidity~\cite{Kaidalov:1979jz}

\begin{equation}\label{etay}
\Delta \eta \approx \ln \left(\frac{M_Y^2}{\langle k_\perp \rangle m_p}\right)\;.
\end{equation}
Taking a sensible value for the transverse momentum $\langle k_\perp \rangle\sim 1$ GeV, and a typical LHC central detector coverage out to $\eta = 5$, i.e. an uninstrumented $\Delta \eta \approx 4-4.5$ for $\sqrt{s}=7-14\,{\rm TeV}$, then gives $M_Y \lesssim 7-10$ GeV. The H1 data therefore corresponds to a very similar region of $M_Y$, and suggests that the admixture of proton--dissociative events at the LHC could be rather large, although to get a more realistic estimate a MC simulation of the system $Y$, including the precise detector acceptances, would clearly have to be performed.

We recall that an analysis of low energy proton--proton data~\cite{Kaydalov:1971ta} suggests that the probability of a low mass, $p\to N^*$ dissociation is about 15\%. In the case of high mass dissociation, we can expect about a 10\% contribution from secondary Reggeons (using the $I\!\!PI\!\!PI\!\!R$ vertex from~\cite{Luna:2008pp}), while the remaining 75\% must then be caused by the triple--Pomeron ($I\!\!PI\!\!PI\!\!P$) term, which is in agreement with the estimates of~\cite{Luna:2008pp}.

In the case of low mass dissociation (i.e. the first 15\% of the proton dissociative contribution) we expect the distribution over the squared momentum transfer, $t$, and the impact parameter, $b_t$, to be more or less the same as in the pure elastic CEP case, as these distributions are driven by the same baryon form factors. On the other hand, the contributions described by the triple--Reggeon terms have a different structure, and are not concentrated in the same regions of $b_t$ space. First, the size of the $I\!\!PI\!\!PI\!\!P$ and the $I\!\!PI\!\!PI\!\!R$ triple--Reggeon vertices are smaller~\cite{Luna:2008pp,Kaidalov:1973tc,Field:1974fg} than the proton size. This is seen for example, in  the H1 measurement of the slope in proton--dissociative events, $b_{\rm pd}=1.79\pm 0.12$ GeV$^{-2}$~\cite{Alexa:2013xxa}, which is lower than that in elastic case, for which  $b_{\rm el}=4.88\pm 0.15\, {\rm GeV}^{-2}$. More importantly than this is the fact that this dissociation is described by the {\em one} proton--
Reggeon vertex, that is by one power of the proton form factor,  $F_p(t)$, while the elastic cross section contains the form factor squared $F_p(t)^2$. In other words, while proton dissociation is described by the first power of the wave function, in the elastic exclusive process we deal with the proton wave function (i.e. parton $b_t$ distribution) {\em squared}, and this contribution is concentrated at much smaller values of $b_t$, where the opacity, and thus the probability of additional inelastic soft interactions, which may fill the rapidity gap, is larger. This was not a problem in the case of photoproduction in $ep$ collisions, studied by H1~\cite{Alexa:2013xxa}, as  here  the effect of absorptive corrections is very small. However the number of  additional (multiple) interactions in high--energy proton--proton collisions is large and we have to account for the role of the gap survival factor, $\langle S^2_{\rm eik}\rangle$ (see for more details, e.g.~\cite{Martin:2009ku,HarlandLang:2011qd,Khoze:
2013dha,Kaidalov:2001iz}) to give a realistic account of the exclusive and dissociative processes. Thus, to translate from the H1 result to the case of proton--proton interactions we have to multiply the high mass contribution to $\sigma_{\rm pd}/\sigma_{\rm el}$ by the ratio of survival factors,
$r_S=\langle S^2_{\rm eik}(\rm pd)\rangle/\langle S^2_{\rm eik}({\rm el})\rangle$. We must also account for the fact that at the LHC the probability of low mass--dissociation is found to be smaller than at low energies: according to the $\sqrt{s}=7$ TeV measurement by the TOTEM~\cite{Antchev:2013haa}, we should instead take a value of 5\% for this contribution. Such a change is expected 
theoretically due to the increased importance of absorptive effects at the higher LHC energies as compared to the fixed--target and ISR data.

To estimate the value of $r_S$ we write the single diffractive cross section, for a pair of mesons of invariant mass $M_X$ produced at central rapidity, $y_X=0$, as 
\begin{equation}\label{cssd}
 \frac{{\rm d}\sigma}{{\rm d}\ln M_Y^2 {\rm d}t} \propto \int \,{\rm d}t^\prime\, \beta(0)\,\beta^2(t)\,g_{3P}(t^\prime)\left(\frac{M_Y^2}{s_0}\right)^{\alpha(0)-1}\left(\frac{s}{M_X^2}\right)^{\alpha(t)-1}\int {\rm d PS}_2\, \frac{|\mathcal{M}^h|^2}{\hat{s}^2}\;,
\end{equation}
where $t$ is the squared momentum transfer to the intact proton, $\beta(t)$ is the Pomeron--nucleon coupling, $\mathcal{M}^h$ is the hard $I\!\!P I\!\!P \to M_3M_4$ amplitude (i.e. as given in (\ref{namp}) with the $s_{13}$, $s_{24}$ terms, and proton form factors $F_p(p_{i\perp}^2)$ removed), which we assume to be point--like and thus have no effect on $r_S$, and $\int {\rm d PS}_2$ indicates the $M_3,M_4$ phase space integration. $t'$ is the squared momentum transfer in the Pomeron loop between $M_Y$ and the hard process $|\mathcal{M}^h|^2$, while we have made the approximation that for centrally produced mesons the cross section scales like $\sim (1/\xi)^{2\alpha(t)-2}$, where $\xi=M_X/\sqrt{s}$ is the momentum fraction transferred through the Pomeron which connects the intact proton to the hard process\footnote{Strictly speaking, this scaling is not consistent with the behaviour of (\ref{namp}), for which e.g. $s_{13}/s_0 \sim M_X \sqrt{s}/s_0$ and not $\sim 1/\xi=\sqrt{s}/M_X$ for centrally produced 
mesons. Rather, we find this behaviour if we make the arguably more physically reasonable assumption that the correct Regge scaling variable is $s_{13}/(|\hat{t}|+\tilde{s}_0)$, where $\hat{t}$ corresponds to the off--shellness of the $t$--channel meson exchange, as in (\ref{namp}), and $\tilde{s}_0$ is an undetermined soft scale. This is reminiscent of the $s/M^2$ scaling we find for high--mass dissociation, and is consistent with the $(\cos\theta_t)^{\alpha(t)}$ behaviour we expect from Regge theory, where $\theta_t$ is the usual $t$--channel scattering angle. However, the effect of including this scaling in (\ref{namp}) can be largely, although not completely, absorbed into a redefinition of the off--shell meson form factor and reasonable choice of $\tilde{s}_0$, and moreover there is a significant uncertainty and freedom in how to include such a $\hat{t}$ dependence when $|\hat{t}| \lesssim \tilde{s}_0$. Nonetheless, measurements of central meson production, with and without proton dissociation may be 
sensitive to such differences. We take this form of the scaling in (\ref{cssd}) because we believe it will give a more accurate prediction for the $M_X$ dependence of $r_S$, although taking the scaling as in ($\ref{namp}$) only leads to a $\sim 20 $\% differences for  experimentally relevant values of $M_X$.}. We use the recent two--channel soft interaction model of~\cite{Khoze:2013dha} together with the $t$--slope measured by H1 in proton--dissociative events\cite{Alexa:2013xxa}. In general the value of $r_S$ will depend on $M_X$, $y_X$ and $M_Y$, due to the non--zero slope of the Reggeon trajectory, which will affect the $b_t$ distribution of the amplitude differently depending on the available rapidity intervals. We therefore present in Table~\ref{trs} estimates for some representative values of $M_X$ and $M_Y$, assuming $y_X=0$ for simplicity. We use model 3 of~\cite{Khoze:2013dha} (which gives an intermediate prediction for the different model choices) for the sake of concreteness and, as well as the 
triple--Reggeon vertex, $G_{3P}=g^1_2$, we allow for multi--Pomeron vertices $g^n_m=G_{3P}(\lambda g_N)^{n+m-3}$, which will provide an additional source of screening corrections in proton--proton interactions, when compared to the HERA case. Taking other versions of the model in~\cite{Khoze:2013dha} gives values which differ by as much as $\pm 20\%$, while excluding multi--Pomeron vertices decreases $r_S$ by $\sim 30 -40\%$. These values also depend on the slope of the Pomeron trajectory, for which we use $\alpha'_P=0.25$ GeV$^{-2}$. 

\begin{table}
\begin{center}
\begin{tabular}{|c|c|c|c|c|}
\hline
$(M_X,M_Y)$ [GeV] &(3,5)&(3,10)&(10,5)&(10,10)\\
\hline
$r_S$&1.68&1.86&1.88&2.08 \\
\hline
\end{tabular}
\caption{Ratios $r_S=\langle S^2_{\rm eik}({\rm pd})\rangle/\langle S^2_{\rm eik}({\rm el})\rangle$ of the soft survival factors for single proton dissociative and pure exclusive production of a state $X$, of mass $M_X$ and rapidity $y_X=0$, and for a mass $M_Y$ of the dissociation system. Values are shown for the model 3 of the soft proton interactions described in~\cite{Khoze:2013dha}, and with multi--Pomeron vertices $g^n_m=G_{3P}(\lambda g_N)^{n+m-3}$ included.}\label{trs}
\end{center}
\end{table}

Thus we can see that while there is some reasonable uncertainty and model--dependence in the precise value, the expected enhancement $r_S$ in the proton dissociative contribution is quite large, with some gentle increase expected at higher $M_X$ and/or $M_Y$.  Very roughly, if we ignore any differences between the mass $M_Y$ probed at HERA and the LHC (which, as discussed above, are expected to be similar), and we take the value $r_S=1.7$ and the HERA measurement of $\sigma_{\rm pd}/\sigma_{\rm el}\sim 0.75$ for high mass dissociation then we expect $\sigma_{\rm pd}(LHC)/\sigma_{\rm el}(LHC)\approx 1.5 \, r_S \approx 2.5$, where we have multiplied by 2 to account for the fact that either proton can dissociate (i.e. assuming the unistrumented regions are symmetric in rapidity, which is not the case at e.g. LHCb), and we have ignored the small contribution from low--mass dissociation for simplicity. While this estimate is clearly quite rough, it demonstrates that the contribution from proton dissociation to 
the current LHC measurements is nonetheless expected to be quite large\footnote{As discussed in~\cite{HarlandLang:2012qz} the addition of Forward Shower Counters\cite{Albrow:2008az,Penzotalk}, which were recently installed by CMS and successfully used in TOTEM + CMS measurements~\cite{CMSeds,Oljemark:2013wsa,Oljemarkeds}, would allow  the contribution from events with comparatively high mass and of a large fraction of events with the low--mass diffractive dissociation to be excluded. The Zero Degree Calorimeters (ZDC), which detects neutral particles in the forward direction, see \cite{Albrow:2008az,Penzotalk,Bell:2012zm}, could also be used during low--luminosity LHC runs to further exclude such dissociative events.}. This contamination is expected to be even larger for processes which are dynamically suppressed in the purely exclusive case by the $J_z^P=0^+$ selection rule: for example, as discussed in~\cite{HarlandLang:2012qz}, the relative contribution from higher--mass dissociation to $\chi_{c(1,2)}$ 
events 
selected with a LRG veto (as in~\cite{LHCbconf}) is expected to be enhanced 
when compared to the $\chi_{c0}$ case due to the $J_z=0$ suppression of the $\chi_{c(1,2)}$ CEP cross sections. However, it is also worth recalling that in the case of higher--mass proton dissociation which, as described above, is expected to give the dominant source of contamination to exclusive events, the momentum transferred to the proton is relatively large, leading to a comparatively  high transverse momentum, $p_{X\perp}$, of the centrally produced system, and this fact can be used to reduce or subtract such dissociative contamination, e.g. by simply placing a cut on higher values of $p_{X\perp}$.

Finally, we recall that at the Tevatron, previous CDF run II studies (see e.g.~\cite{Aaltonen:2011hi,Albrow:2009nj} for earlier references), for which the events were also selected using a LRG veto, had a nearly full rapidity coverage, and so the contribution from events with unseen proton dissociation was practically negligible. In the new measurement~\cite{Albrow:2013mva,Mikeeds} of central $\pi^+\pi^-$ production, where the rapidity coverage is not quite as extensive, the contribution from such events may be somewhat more important, although should still be quite small. In particular the unistrumented rapidity region ranges from $\Delta \eta \approx 1 - 1.7$ for $\sqrt{s}=900 - 1960$ GeV, corresponding to $M_Y \lesssim 1.5 - 2.2$ GeV. Such low mass dissociation should only give a small $\sim 20\%$ contribution to the purely exclusive events. Indeed, for $\sqrt{s}=900$ GeV the rapidity interval is particularly small, and so such dissociation should be nearly absent.

\section{Conclusion}\label{conc}

In this paper we have presented a detailed analysis of central exclusive meson pair production within the framework of Regge theory, as depicted in Fig.~\ref{npip}. Such an approach is expected to be relevant at lower values of the meson transverse momentum $k_\perp$ and/or pair invariant mass $M_X$, and may be particularly important for the case of flavour--non--singlet mesons ($\pi\pi$, $KK$...), for which the perturbative contribution is expected to be dynamically suppressed, see~\cite{HarlandLang:2011qd}. We have provided a detailed description of a phenomenological model for such processes, which on the one hand applies the well--established tools of Regge theory, but on the other is still currently somewhat unconstrained in its key ingredients. Such a model compares well to the existing ISR data on exclusive $\pi^+\pi^-$ and $K^+K^-$ production~\cite{Breakstone:1990at,Breakstone:1989ty}, but the new preliminary CDF data on $\pi^+\pi^-$ CEP at $\sqrt{s}$ = 900 and 1960 GeV, presented in~\cite{Albrow:
2013mva,Mikeeds}, as well as the forthcoming data from CMS~\cite{enterria}, CMS+Totem~\cite{CMSeds,Oljemark:2013wsa,Oljemarkeds}, ATLAS+ALFA~\cite{Staszewski:2011bg,Sykoraeds}, RHIC~\cite{Leszek} and LHCb~\cite{Paula} represent a new and potentially extensive test of this approach. This continuum production process also represents an irreducible background to the CEP of resonant states ($f_0, f_2, \chi_{c(0,2)}$...) via two--body decays to mesons.

Motivated by this, in this paper we have implemented this phenomenological model in the new public \texttt{Dime} Monte Carlo~\cite{dime} for meson pair ($\pi\pi, KK, \rho\rho$) CEP. We give the user freedom to set the most important, and not fully constrained, aspects and parameters of the model, so that these can be compared with and constrained by future data. We also include the soft survival factor at the fully differential level, which (as described in e.g.~\cite{HarlandLang:2010ep}) is crucial to give a complete prediction, in particular when considering the distribution of the outgoing intact protons.

In Section~\ref{numer} we have used this MC to make detailed numerical predictions for the Tevatron and LHC, demonstrating how different observables may be used to further test and constrain the phenomenological model. We have also shown that the distribution in azimuthal angle between the outgoing protons is highly sensitive to soft survival effects, with striking `diffractive dips' appearing when various cuts are placed on the proton $p_\perp$. In this way, measurements of exclusive meson pair production with tagged protons may be used as a novel probe of the models of hadronic interactions used to calculate the soft survival factors. Such measurements are possible at the LHC, with the CMS+Totem~\cite{CMSeds,Oljemark:2013wsa,Oljemarkeds} and ATLAS+ALFA~\cite{Staszewski:2011bg,Sykoraeds} during special low luminosity running conditions, and are already begin made at RHIC by the STAR collaboration~\cite{Leszek}.

In~\cite{HarlandLang:2011qd,Harland-Lang:2013ncy} it was shown how meson pair CEP may be modelled in a pQCD based framework, which should be relevant at sufficiently high meson transverse momentum $k_\perp$. Such an approach leads to many non--trivial predictions and displays some remarkable theoretical features, as summarized in~\cite{Harland-Lang:2013qia}. However, as the meson $k_\perp$ decreases, we would not expect to trust such an approach, and so we must instead consider a more model--dependent formalism, as described in this paper. Nonetheless, this model, while not firmly grounded in QCD, still presents an interesting and rich phenomenology, which is beginning to be explored with new analyses from the Tevatron and forthcoming data from the LHC and RHIC.  More generally, we may hope in the future to experimentally probe the transition between these two regimes, an issue which still remains unclear. In the context of the model discussed in this paper, this transition is highly sensitive to  the form 
factor for the coupling of Pomeron to the meson pair production subprocess. We have seen how the preliminary CDF measurement~\cite{Albrow:2013mva,Mikeeds} of $\pi^+\pi^-$ production seems to be described better by an `Orear' type form factor ($\sim \exp(-b k_\perp)$). On the other hand we have also shown that such a form factor, which falls relatively gently with the meson $k_\perp$ in comparison to the standard `soft' exponential behaviour ($\sim \exp(-b k_\perp^2)$), tends to predict a $\pi^0\pi^0$ cross section at higher $k_\perp$ that is in strong conflict with the CDF measurement of $\gamma\gamma$ CEP~\cite{Aaltonen:2011hi}, and corresponding limit on $\pi^0\pi^0$ production. Further measurements will therefore be crucial in further clarifying this uncertain question.

The central exclusive production of meson pairs therefore represents a process of much phenomenological interest, which can shed light on both perturbative and non--perturbative aspects of QCD. Moreover it is of particular experimental relevance, with a range of forthcoming and existing hadron collider data to consider. In this paper we have provided the tools for a more in depth comparison of the existing theory with such data, and in this way to shed further light on this interesting process.

\section*{Acknowledgements}
We thank Mike Albrow, Paula Collins, David d'Enterria, Jonathan  Hollar, Risto Orava, Kenneth Osterberg, James Stirling and Antoni Szczurek for useful discussions. This work was supported by the grant RFBR 14-02-00004 and by the Federal Program of the Russian State RSGSS-4801.2012.2. MGR thank the IPPP at the University of Durham for hospitality. VAK thanks the Kavli Institute for Theoretical Physics at USCB for hospitality during the preparation of this work.

\bibliography{ggbib}{}

\begin{thebibliography}{10}

\bibitem{Martin:2009ku}
A.~D. Martin, M.~G. Ryskin, and V.~A. Khoze,
\newblock Acta Phys.Polon. {\bf B40}, 1841 (2009), 0903.2980.

\bibitem{Albrow:2010yb}
M.~G. Albrow, T.~D. Coughlin, and J.~R. Forshaw,
\newblock Prog.Part.Nucl.Phys. {\bf 65}, 149 (2010), 1006.1289.

\bibitem{HarlandLang:2013jf}
L.~A. Harland-Lang, V.~A. Khoze, M.~G. Ryskin, and W.~J. Stirling,
\newblock (2013), 1301.2552.

\bibitem{HarlandLang:2011qd}
L.~A. Harland-Lang, V.~A. Khoze, M.~G. Ryskin, and W.~J. Stirling,
\newblock Eur.Phys.J. {\bf C71}, 1714 (2011), 1105.1626.

\bibitem{Harland-Lang:2013ncy}
L.~A. Harland-Lang, V.~A. Khoze, M.~G. Ryskin, and W.~J. Stirling,
\newblock Eur.Phys.J. {\bf C73}, 2429 (2013), 1302.2004.

\bibitem{Harland-Lang:2013qia}
L.~A. Harland-Lang, V.~A. Khoze, M.~G. Ryskin, and W.~J. Stirling,
\newblock Phys.Lett. {\bf B725}, 316 (2013), 1304.4262.

\bibitem{HarlandLang:2010ep}
L.~A. Harland-Lang, V.~A. Khoze, M.~G. Ryskin, and W.~J. Stirling,
\newblock Eur.Phys.J. {\bf C69}, 179 (2010), 1005.0695.

\bibitem{Brodsky:1981rp}
S.~J. Brodsky and G.~P. Lepage,
\newblock Phys.Rev. {\bf D24}, 1808 (1981).

\bibitem{Benayoun:1989ng}
M.~Benayoun and V.~L. Chernyak,
\newblock Nucl.Phys. {\bf B329}, 285 (1990).

\bibitem{Kaidalov:1974qi}
A.~Kaidalov and K.~Ter-Martirosyan,
\newblock Nucl.Phys. {\bf B75}, 471 (1974).

\bibitem{Azimov:1974fa}
Y.~I. Azimov, V.~A. Khoze, E.~M. Levin, and M.~G. Ryskin,
\newblock Sov.J.Nucl.Phys. {\bf 21}, 215 (1975).

\bibitem{Pumplin:1976dm}
J.~Pumplin and F.~Henyey,
\newblock Nucl.Phys. {\bf B117}, 377 (1976).

\bibitem{Desai:1978rh}
B.~R. Desai, B.~C. Shen, and M.~Jacob,
\newblock Nucl.Phys. {\bf B142}, 258 (1978).

\bibitem{HarlandLang:2012qz}
L.~A. Harland-Lang, V.~A. Khoze, M.~G. Ryskin, and W.~J. Stirling,
\newblock Eur.Phys.J. {\bf C72}, 2110 (2012), 1204.4803.

\bibitem{Lebiedowicz:2009pj}
P.~Lebiedowicz and A.~Szczurek,
\newblock Phys.Rev. {\bf D81}, 036003 (2010), 0912.0190.

\bibitem{Lebiedowicz:2012nk}
P.~Lebiedowicz and A.~Szczurek,
\newblock (2012), 1212.0166.

\bibitem{Collins:1977jyp}
P. D. B. Collins, \emph{An Introduction to Regge Theory and High-Energy
  Physics} (1977).

\bibitem{Lebiedowicz:2013ika}
P.~Lebiedowicz, O.~Nachtmann, and A.~Szczurek,
\newblock (2013), 1309.3913.

\bibitem{enterria}
David d'Enterria, private communication.

\bibitem{CMSeds}
Christina Mesropian (CMS Collaboration) , talk at EDS Blois 2013 Workshop,
  Saariselka, Lapland, September 9-13.

\bibitem{Oljemark:2013wsa}
TOTEM collaboration, F.~Oljemark,
\newblock (2013), 1310.4305.

\bibitem{Oljemarkeds}
Fredrik Oljemark, talk at EDS Blois 2013 Workshop, Saariselka, Lapland,
  September 9-13.

\bibitem{Staszewski:2011bg}
R.~Staszewski, P.~Lebiedowicz, M.~Trzebinski, J.~Chwastowski, and A.~Szczurek,
\newblock Acta Phys.Polon. {\bf B42}, 1861 (2011), 1104.3568.

\bibitem{Sykoraeds}
Tomas Sykora, talk at EDS Blois 2013 Workshop, Saariselka, Lapland, September
  9-13.

\bibitem{Leszek}
Leszek Adamczyk, talk at the Workshop 15th conference on Elastic and
  Diffractive scattering, EDS Blois 2013 Workshop, Saariselka, Lapland,
  September 9-13.

\bibitem{Paula}
Paula Collins, talk at `Results and Prospects of Forward Physics at the LHC',
  CERN, Feb. 11-13, 2013.

\bibitem{Albrow:2013mva}
M.~Albrow, A.~Swiech, and M.~Zurek,
\newblock (2013), 1310.3839.

\bibitem{Mikeeds}
Mike Albrow, talk at EDS Blois 2013 Workshop, Saariselka, Lapland, September
  9-13.

\bibitem{Ryskin:2009tk}
M.~G. Ryskin, A.~D. Martin, and V.~A. Khoze,
\newblock Eur.Phys.J. {\bf C60}, 265 (2009), 0812.2413.

\bibitem{Khoze:2013dha}
V.~A. Khoze, A.~D. Martin, and M.~G. Ryskin,
\newblock Eur.Phys.J. {\bf C73}, 2503 (2013), 1306.2149.

\bibitem{Khoze:2013jsa}
V.~Khoze, A.~Martin, and M.~Ryskin,
\newblock Eur.Phys.J. {\bf C74}, 2756 (2014), 1312.3851.

\bibitem{Gotsman:2012rm}
E.~Gotsman, E.~Levin, and U.~Maor,
\newblock Phys.Lett. {\bf B716}, 425 (2012), 1208.0898.

\bibitem{Gotsman:2012rq}
E.~Gotsman, E.~Levin, and U.~Maor,
\newblock Phys.Rev. {\bf D85}, 094007 (2012), 1203.2419.

\bibitem{fortheCOMPASS:2013vda}
Alexander Austregesilo for the COMPASS Collaboration,
\newblock (2013), 1310.3190.

\bibitem{dime}
The Dime code and documentation are available at {\tt
  http://dimemc.hepforge.org/}.

\bibitem{Lebiedowicz:2011nb}
P.~Lebiedowicz, R.~Pasechnik, and A.~Szczurek,
\newblock Phys.Lett. {\bf B701}, 434 (2011), 1103.5642.

\bibitem{Donnachie:1992ny}
A.~Donnachie and P.~V. Landshoff,
\newblock Phys.Lett. {\bf B296}, 227 (1992), hep-ph/9209205.

\bibitem{Orear:1964zz}
J.~Orear,
\newblock Phys.Rev.Lett. {\bf 12}, 112 (1964).

\bibitem{Anselm1967479}
A.~Anselm and I.~Dyatlov,
\newblock Physics Letters B {\bf 24}, 479  (1967).

\bibitem{Andreev:1967zz}
I.~Andreev and I.~M. Dremin,
\newblock Pisma Zh.Eksp.Teor.Fiz. {\bf 6}, 810 (1967).

\bibitem{Mandelstam:1963cw}
S.~Mandelstam,
\newblock Nuovo Cim. {\bf 30}, 1148 (1963).

\bibitem{Amati:1962nv}
D.~Amati, A.~Stanghellini, and S.~Fubini,
\newblock Nuovo Cim. {\bf 26}, 896 (1962).

\bibitem{Yu:2011zu}
B.~G. Yu, T.~K. Choi, and W.~Kim,
\newblock Phys.Rev. {\bf C83}, 025208 (2011), 1103.1203.

\bibitem{Khoze:2002nf}
V.~A. Khoze, A.~D. Martin, and M.~G. Ryskin,
\newblock Eur.Phys.J. {\bf C24}, 581 (2002), hep-ph/0203122.

\bibitem{Ryskin:2009tj}
M.~G. Ryskin, A.~Martin, and V.~A. Khoze,
\newblock Eur.Phys.J. {\bf C60}, 249 (2009), 0812.2407.

\bibitem{Ryskin:2011qe}
M.~G. Ryskin, A.~D. Martin, and V.~A. Khoze,
\newblock Eur.Phys.J. {\bf C71}, 1617 (2011), 1102.2844.

\bibitem{Khoze:2014aca}
V.~Khoze, A.~Martin, and M.~Ryskin,
\newblock (2014), 1402.2778.

\bibitem{Khoze:2000cy}
V.~A. Khoze, A.~D. Martin, and M.~G. Ryskin,
\newblock Eur.Phys.J. {\bf C14}, 525 (2000), hep-ph/0002072.

\bibitem{HarlandLang:2010ys}
L.~A. Harland-Lang, V.~A. Khoze, M.~G. Ryskin, and W.~J. Stirling,
\newblock Eur.Phys.J. {\bf C71}, 1545 (2011), 1011.0680.

\bibitem{Harland-Lang:2013bya}
L.~A. Harland-Lang, V.~A. Khoze, and M.~G. Ryskin,
\newblock (2013), 1310.2759.

\bibitem{Breakstone:1990at}
Ames-Bologna-CERN-Dortmund-Heidelberg-Warsaw Collaboration, A.~Breakstone {\em
  et~al.},
\newblock Z.Phys. {\bf C48}, 569 (1990).

\bibitem{Breakstone:1989ty}
Ames-Bologna-CERN-Dortmund-Heidelberg-Warsaw Collaboration, A.~Breakstone {\em
  et~al.},
\newblock Z.Phys. {\bf C42}, 387 (1989).

\bibitem{Aaltonen:2011hi}
CDF Collaboration, T.~Aaltonen {\em et~al.},
\newblock Phys.Rev.Lett. {\bf 108}, 081801 (2012), 1112.0858.

\bibitem{Aaltonen:2009kg}
CDF Collaboration, T.~Aaltonen {\em et~al.},
\newblock Phys.Rev.Lett. {\bf 102}, 242001 (2009), 0902.1271.

\bibitem{LHCbconf}
LHCb collaboration, CERN-LHCb-CONF-2011-022.

\bibitem{Alexa:2013xxa}
H1 Collaboration, C.~Alexa {\em et~al.},
\newblock Eur.Phys.J. {\bf C73}, 2466 (2013), 1304.5162.

\bibitem{Kaidalov:1979jz}
A.~B. Kaidalov,
\newblock Phys.Rept. {\bf 50}, 157 (1979).

\bibitem{Kaydalov:1971ta}
A.~B. Kaidalov,
\newblock Sov.J.Nucl.Phys. {\bf 13}, 226 (1971).

\bibitem{Luna:2008pp}
E.~G.~S. Luna, V.~A. Khoze, A.~D. Martin, and M.~G. Ryskin,
\newblock Eur.Phys.J. {\bf C59}, 1 (2009), 0807.4115.

\bibitem{Kaidalov:1973tc}
A.~B. Kaidalov, V.~A. Khoze, Y.~F. Pirogov, and N.~L. Ter-Isaakyan,
\newblock Phys.Lett. {\bf B45}, 493 (1973).

\bibitem{Field:1974fg}
R.~Field and G.~Fox,
\newblock Nucl.Phys. {\bf B80}, 367 (1974).

\bibitem{Kaidalov:2001iz}
A.~Kaidalov, V.~A. Khoze, A.~D. Martin, and M.~Ryskin,
\newblock Eur.Phys.J. {\bf C21}, 521 (2001), hep-ph/0105145.

\bibitem{Antchev:2013haa}
TOTEM, G.~Antchev {\em et~al.},
\newblock Europhys.Lett. {\bf 101}, 21003 (2013).

\bibitem{Albrow:2008az}
M.~Albrow {\em et~al.},
\newblock JINST {\bf 4}, P10001 (2009), 0811.0120.

\bibitem{Penzotalk}
Aldo Penzo, talk at Forward Physics at the LHC meeting, Reggio Calabria,15-18
  July 2013.

\bibitem{Bell:2012zm}
A.~J. Bell {\em et~al.},
\newblock (2012), FERMILAB-PUB-10-718-CMS, CERN-CMS-NOTE-2010-015.

\bibitem{Albrow:2009nj}
CDF Collaboration, M.~Albrow,
\newblock (2009), 0909.3471.

\end{thebibliography}
\bibliographystyle{h-physrev}

\end{document}